\documentclass{aa}  

\usepackage[utf8]{inputenc}

\usepackage{amsmath,amsfonts,amssymb}
\usepackage{upgreek}
\usepackage{graphicx}
\usepackage{txfonts}
\usepackage[colorlinks=true, allcolors=blue]{hyperref}
\usepackage{natbib}
\usepackage{widetext}
\usepackage{svg}
\usepackage{textgreek}
\usepackage{dblfloatfix}
\usepackage{soul}
\usepackage{makecell}

\usepackage{xcolor}
\newcommand{\manon}[1]{\textcolor{black}{#1}} 
 
\newcommand{\rev}[1]{\textcolor{black}{#1}}
\newcommand{\revv}[1]{\textcolor{black}{#1}}
\newcommand{\corr}[1]{\textcolor{black}{#1}}

\title{Spectroscopy using a visible photonic lantern at the Subaru Telescope: Laboratory characterization and the first on-sky demonstration on Ikiiki ($\alpha$ Leo) and `Aua ($\alpha$ Ori)}

\titlerunning{Spectroscopy with a Visible Photonic Lantern}

\author{%
S.~Vievard \inst{\ref{a},\ref{b}} \and %
M.~Lallement\inst{\ref{a},\ref{c}}  \and %
S.~Leon-Saval\inst{\ref{d}} \and %
O.~Guyon\inst{\ref{a},\ref{b},\ref{e},\ref{f}} \and %
N.~Jovanovic\inst{\ref{g}} \and
E.~Huby\inst{\ref{c}} \and %
S.~Lacour\inst{\ref{c}} \and
J.~Lozi\inst{\ref{a}} \and
V.~Deo\inst{\ref{a}} \and
K.~Ahn\inst{\ref{a},\ref{kasi}} \and
M.~Lucas\inst{\ref{k}} \and
S.~Sallum\inst{\ref{ucsi}} \and
B.~Norris\inst{\ref{d},\ref{h},\ref{i}} \and
C.~Betters\inst{\ref{d}} \and %
R.~Amezcua-Correa\inst{\ref{j}} \and
S.~Yerolatsitis\inst{\ref{j}} \and
M.~P.~Fitzgerald\inst{\ref{ucla}} \and
J.~Lin\inst{\ref{ucla}} \and
Y.J.~Kim\inst{\ref{ucla}} \and
P.~Gatkine\inst{\ref{ucla}} \and
T.~Kotani\inst{\ref{b}} \and
M.~Tamura\inst{\ref{b},\ref{univT},\ref{naoj}} \and 
T.~Currie \inst{\ref{a},\ref{UTexas}} \and %
H-D.~Kenchington\inst{\ref{c}} \and 
G.~Martin\inst{\ref{l}} \and
G.~Perrin\inst{\ref{c}} }

\institute{National Astronomical Observatory of Japan, Subaru Telescope, 650 North Aohoku Place, Hilo, HI 96720, U.S.A. email: vievard@naoj.org \label{a}
\and Astrobiology Center, 2-21-1, Osawa, Mitaka, Tokyo, 181-8588, Japan \label{b} 
\and LESIA, Observatoire de Paris, Universite PSL, CNRS, Sorbonne Universite, Sorbonne Paris Cite, 5 place Jules Janssen, 92195 Meudon, France  \label{c}
\and Sydney Astrophotonic Instrumentation Laboratory, School of Physics, The University of Sydney, Sydney, NSW 2006, Australia \label{d}
\and Steward Observatory, University of Arizona, Tucson, AZ 85721, USA \label{e}
\and College of Optical Sciences, University of Arizona, Tucson, AZ 85721, U.S.A. \label{f} 
\and California Institute of Technology, 1200 E California Blvd, Pasadena, CA 91125, U.S.A. \label{g} 
\and Korea Astronomy and Space Science Institute (KASI), Daejeon 34055, Republic of Korea \label{kasi}
\and Univ. of California, Irvine, G302 C Student Center, Irvine, CA 92697 \label{ucsi}
\and Univ. of California, Los Angeles, 405 Hilgard Avenue, CA 90095 \label{ucla}
\and Sydney Institute for Astronomy, School of Physics, The University of Sydney, NSW 2006, Australia \label{h} 
\and AAO-USyd, School of Physics, University of Sydney 2006 \label{i}
\and The College of Optics and Photonics, University of Central Florida, 4304 Scorpius St, Orlando, FL 32816 \label {j}
\and Institute for Astronomy, University of Hawai'i, 640 N. Aohoku Pl, Hilo, HI 96720 \label{k}
\and The University of Tokyo, 7-3-1 Hongo, Bunkyo-ku, Tokyo 113-0033, Japan \label{univT}
\and NAOJ, 2-21-1 Osawa, Mitaka, Tokyo 181-8588, Japan \label{naoj}
\and Department of Physics and Astronomy, University of Texas at San Antonio, San Antonio, TX 78006, USA \label{UTexas}
\and Univ. Grenoble Alpes, CNRS, IPAG, 414 Rue de la Piscine, 38400 Saint-Martin-d'Hères, France \label{l}}

\abstract
{Photonic lanterns (PLs) are waveguide devices enabling high-throughput single-mode spectroscopy and high angular resolution.}
{We aim to present the first on-sky demonstration of a PL operating in visible light, to measure its throughput and assess its potential for high-resolution spectroscopy of compact objects.}
{We used the SCExAO instrument (a double-stage extreme adaptive optics system installed at the Subaru Telescope) and FIRST mid-resolution spectrograph (R~3000) to test the visible capabilities of the PL on internal source and on-sky observations.} 
{The best averaged coupling efficiency over the PL field of view was measured at $51\% \pm 10\%$, with a peak at $80\%$. We also investigated the relationship between coupling efficiency and the Strehl ratio for a PL, comparing them with those of a single-mode fiber (SMF). Findings show that in the adaptive optics regime a PL offers a better coupling efficiency performance than an SMF, especially in the presence of low-spatial-frequency aberrations. 
We observed Ikiiki ($\alpha$ Leo - $m_R$ = 1.37) and `Aua ($\alpha$ Ori - $m_R$ = -1.17) at a frame rate of 200 Hz. Under median seeing conditions (about 1~arcsec measured in the H band) and large tip or tilt residuals (over 20~mas), we estimated an average light coupling efficiency of \revv{$14.5\% \pm 7.4\%$, with a maximum of $42.8\%$ at 680~nm}. We were able to reconstruct both star's spectra, containing various absorption lines.}
{The successful demonstration of this device opens new possibilities in terms of high-throughput single-mode fiber-fed spectroscopy in the visible. The demonstrated on-sky coupling efficiency performance would not have been achievable with a single SMF injection setup under similar conditions, partly because the residual tip or tilt alone exceeded the field of view of a visible SMF (18 mas at 700 nm). This emphasizes the enhanced resilience of PL technology to such atmospheric disturbances. The additional capabilities in high angular resolution are also promising but still have to be demonstrated in a forthcoming investigation.}{}

\begin{document} 
\maketitle
\section{Introduction}

Spectroscopy at high throughput and high resolution plays a key role in the study of the Universe, whether it is used for exoplanet detection or characterization, stellar physics, or cosmology. 
High-dispersion spectroscopic (HDS) observations of star-planet systems have recently been used to isolate giant exoplanet signals via Doppler shifts, enabling measurements of orbital velocities, masses, and molecular features~\citep[e.g.,][]{snellen10,brogi12,currie2023ppvii}.
This HDS analysis approach, coupled with high contrast imaging, is also expected to enable rocky planet characterization with next-generation telescopes~\citep{snellen15}.
High-spectral-resolution observations of stars have enabled a broad range of science, from measurements of circumstellar disk and wind kinematics~\citep[e.g.,][]{pontoppidan11} and ices~\citep[e.g.,][]{boogert02}, to tests of general relativity via radial velocity measurements of stars in the galactic center~\citep[e.g.,][]{do19}.
High-throughput, high-resolution spectrographs are also crucial for a wide range of extragalactic faint source applications, from transient follow-up~\citep[e.g.,][]{patat07}, to characterizing stellar populations~\citep[e.g.,][]{reddy22}, to searches for intermediate-mass black holes~\citep[e.g.,][and references therein]{casares14}. 

These broad and impactful applications motivate the development of technologies that can maximize throughput, while preserving high spectral resolution and quality. 
Compared to traditional multimode (MM) or slit-based spectroscopy, single-mode fiber (SMF) fed spectroscopy offers numerous advantages for astronomical observations, ranging from improved spectral quality and resolution to increased instrument compactness and stability. \rev{The beam coming out of an SMF is indeed highly stable in shape, since it is determined by the fundamental mode of the fiber, ensuring a stable spectral response of the spectrograph.} The weakness of SMFs lies in the challenge of light injection, requiring very fine alignment of the source to the fiber core and a high Strehl ratio, which are difficult to obtain, especially in visible light. Light from a telescope can be efficiently injected into an SMF with the help of extreme adaptive optics (ExAO) \citep{jovanovic2017efficient}. 
The ExAO systems are designed to measure and correct wavefront aberrations induced by atmospheric turbulence, enabling Strehl ratios up to $95\%$ in the $H$ band. 

Multiple post-ExAO SMF-based instruments have been or will be commissioned on major observatories: Subaru/REACH~\citep{kotani2020reach}, Keck/KPIC~\citep{delorme2021keck}, Subaru/GLINT~\citep{martinod2021scalable}, VLT/HiRISE~\citep{vigan2024first}, and Subaru/FIRST~\citep{vievard2023singleaperture}. While some of them differ by their wavelength operation (Subaru/FIRST is in the visible, the rest are in the near-infrared), all of them share the same struggle of coupling efficiency and throughput limitations.

A solution that combines high coupling efficiency with the stability and compactness advantages of SMF-based spectroscopy was proposed by \cite{leon2005multimode}. It consists of using a device transitioning from an MM input to multiple single-mode (SM) outputs. The MM input ensures high coupling efficiency, the slow transition between MM and SM ensures high throughput, and the SM outputs enable the use of SMF-fed spectroscopy. Such devices are called photonic lanterns (PLs). The interest in PLs also lies in their sensitivity to the input scene and wavefront: the flux distribution among the PL outputs depends both on the source position and shape~\citep{kim2022spectroastrometry} and on the wavefront input~\citep{norris2020all}. The corresponding information allows for retrieval of the geometry and position of the observed object, as well as wavefront sensing and optimization of injection efficiency into one specific output SMF if desired~\citep{Norris2022}. These features are not possible with SMFs; hence, PLs can potentially provide access to a wider range of science applications.

The characteristics described above make PLs an attractive technology for high-contrast imaging science.
One particular example is spatially and spectrally resolved searches for accreting protoplanets orbiting nearby young stars.
Conventional imaging of these targets is typically limited to $\sim150-200$ mas angular separations \citep{keppler2018,haffert19,currie2022}, corresponding to an orbital separation of $\gtrsim20$~AU  for targets at $\sim$150 pc. 
Simulations of both spectroastrometry and interferometric imaging with PLs have shown that they can be used to resolve a few AU orbits around nearby young stars~\citep{xin2022efficient,kim2022spectroastrometry,levinstein23,kim24interferometry} -- a region where the majority of known mature giant planets reside~\citep{fernandes19}.
Coupling the SMF outputs to a medium-resolution spectrograph also allows for spectral-differential measurements that can isolate accretion-tracing hydrogen lines~\citep{kim2022spectroastrometry,levinstein23}.
These excesses, along with spectral slopes, can distinguish between scattered light from circumstellar disk material and self-luminous planets~\citep[e.g.,][]{haffert19,currie2022,wagner23}, which is crucial for bona fide protoplanet detections.
Beyond protoplanets, PLs could pay dividends for a broad range of science cases requiring high-angular-resolution observations, where interferometry is often utilized, including studies of close-in spirals in protoplanetary disks~\citep[e.g.,][]{sallum19}, inner regions of active galactic nucleus tori~\citep[e.g.,][]{ford14}, and Solar System object surfaces~\citep[e.g.,][]{conrad15}.


The first on-sky demonstration of a PL was performed on the 3.9~m Anglo-Australian Telescope~\citep{cvetojevic2012first}. The PL was coupled to an arrayed waveguide grating (AWG) spectrograph used in tandem with the IRIS2 imaging spectrograph~\citep{tinney2004iris2}, providing a resolution power of about 2500. The interfacing on the Cassegrain focus of the telescope and the absence of wavefront correction by an adaptive system limited the  coupling to about $5$ to $7\%$ depending on the seeing variations.

We recently acquired a PL with 19~SMF outputs, designed for visible wavelengths. The device was installed at the 8.2~m Subaru Telescope as a sub-module of the Subaru Extreme Adaptive Optics (SCExAO) platform \rev{to explore the possibilities of using PLs for high-throughput and high-resolution spectroscopy.} \rev{Stellar light can be injected into the PL in a pupil plane or in a focal plane. This paper focuses on the latter approach, though the former is also a possibility.} The 19 PL outputs feed the Subaru/FIRST mid-resolution spectrograph (R$\sim3000$) optimized for wavelengths from 600 to 800~nm~\citep{vievard2023photonic}. An SMF is also installed in parallel and feeds the same spectrograph for a comparison between the two methods. 

In Sect.~\ref{PL-scexao}, we provide a comprehensive description of the PL device, and its integration on SCExAO. Section~\ref{PL-inlab} presents an in-lab characterization of light injection within the PL. Finally, Sect.~\ref{PL-onsky} presents the first demonstration of spectroscopy in visible wavelengths, performed using an ExAO-fed PL. We observe and reconstruct the spectrum of two targets:  Ikiiki~\footnote{This paper refers to the main targets per their Hawaiian names, in honor of Mauna Kea, the mountain from which we were most fortunate to perform our observations, and with acknowledgment of the indigenous hawaiian communities who have had and continue to have Kuleana (translated into English as "responsibility") to this land.} ($\alpha$ Leo) and `Aua ($\alpha$ Ori).


\section{A photonic lantern spectrograph at the Subaru Telescope}
\label{PL-scexao}


\subsection{The 19~pigtailed outputs Photonic Lantern}
\begin{figure}[t]
\centering
        \includegraphics[width=\linewidth]{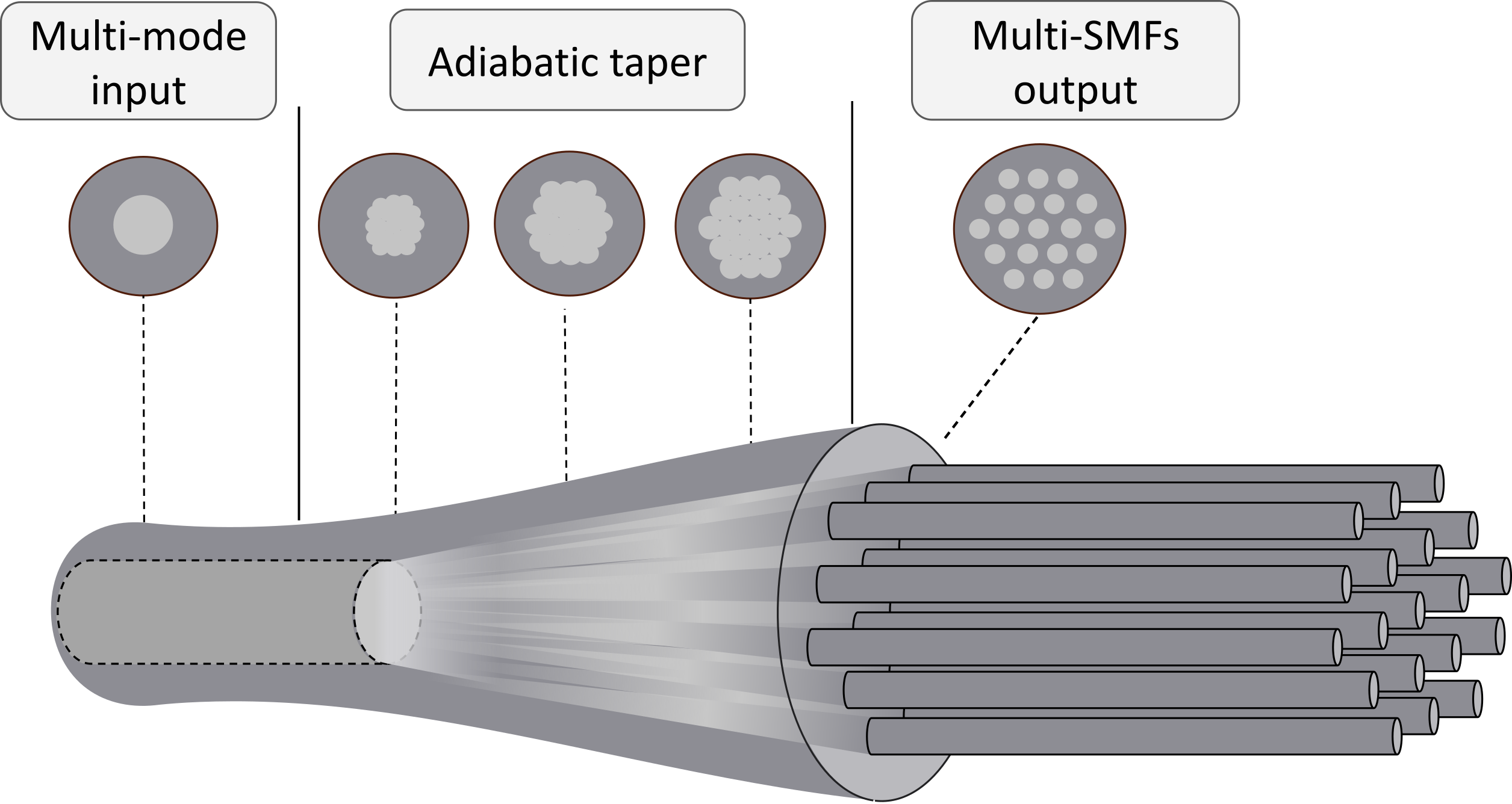}
        \caption{Photonic lantern principle: light is injected in the MM input. The waveguide adiabatically transitions to several SMFs.}
        \label{fig:PL_principle}
\end{figure}

The PL used in this experiment has 1 MM input and 19~SM outputs. The SM outputs of a PL come in various configurations~\citep{leon2010photonic}. Our lantern was constructed from separate SMFs where the outputs are separate pigtails (see the drawing in Fig.~\ref{fig:PL_principle}). An adiabatic taper allows for an efficient and low-loss mode conversion (throughput greater than 90\%;~\citealp{birks2015photonic}). 
This was accomplished by placing the multiple SM fibers into a low refractive-index capillary (that has an index less than that of the SM fiber cladding, which in turn is less than that of the SM fiber core). This assembly is tapered down, such that the cores of the SM fibers disappear, their cladding becomes the new core of a MM region, and the capillary becomes the MM region's cladding.
This smooth transition is crucial, since abrupt changes in the waveguide geometry can lead to internal reflections and mode mismatch, which would result in decreasing the device throughput. In the transition region, light propagating in the modes supported by the input waveguide section is redistributed across the modes of the output - this process is reciprocal. As long as the number of output SM fibers is equal to or exceeds the number of unique modes supported in the input MM region, this process has very high efficiency. \rev{The PL used in this experiment was designed to work at the lower end of the FIRST wavelength coverage (around 650~nm). This design choice ensures that, at the shortest wavelengths, the number of modes does not increase significantly, which would otherwise contribute to transmission losses in the PL device.} The 19~SMFs were then spliced to a linear \corr{V-groove (VG)}, allowing a spectrograph to cross-disperse the light.

{}

\begin{figure}[t]
\centering
        \includegraphics[width=\linewidth]{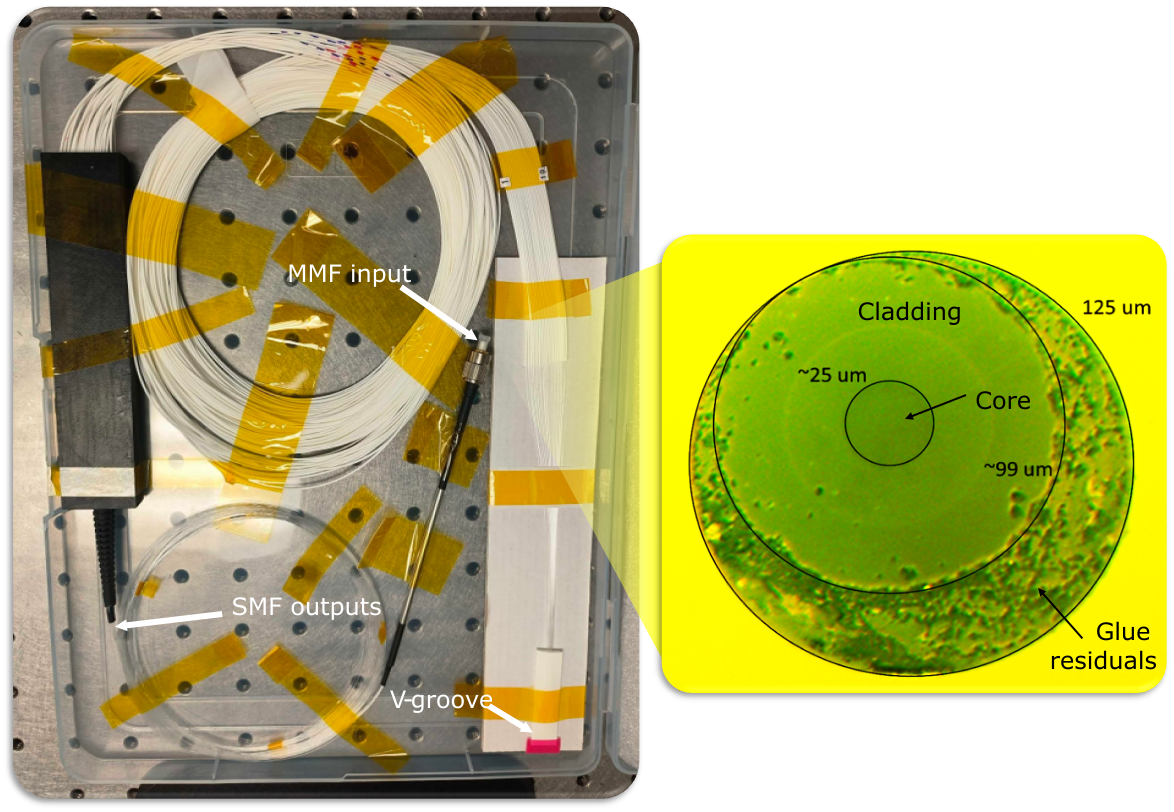}
        \caption{Photonic lantern hardware. The input is an MM fiber, and the outputs are SMFs spliced to a VG.}
        \label{fig:pl-input}
\end{figure}

\rev{Figure~\ref{fig:pl-input} shows the final device, spliced to the V-groove. We also provide a picture of the PL input taken thanks to a microscope, where we can distinguish the fiber core and the cladding whose dimensions are respectively about $25~\mu$m and $99~\mu$m diameter.}

{}

\begin{figure*}[!h]
\centering
    \includegraphics[width=\linewidth]{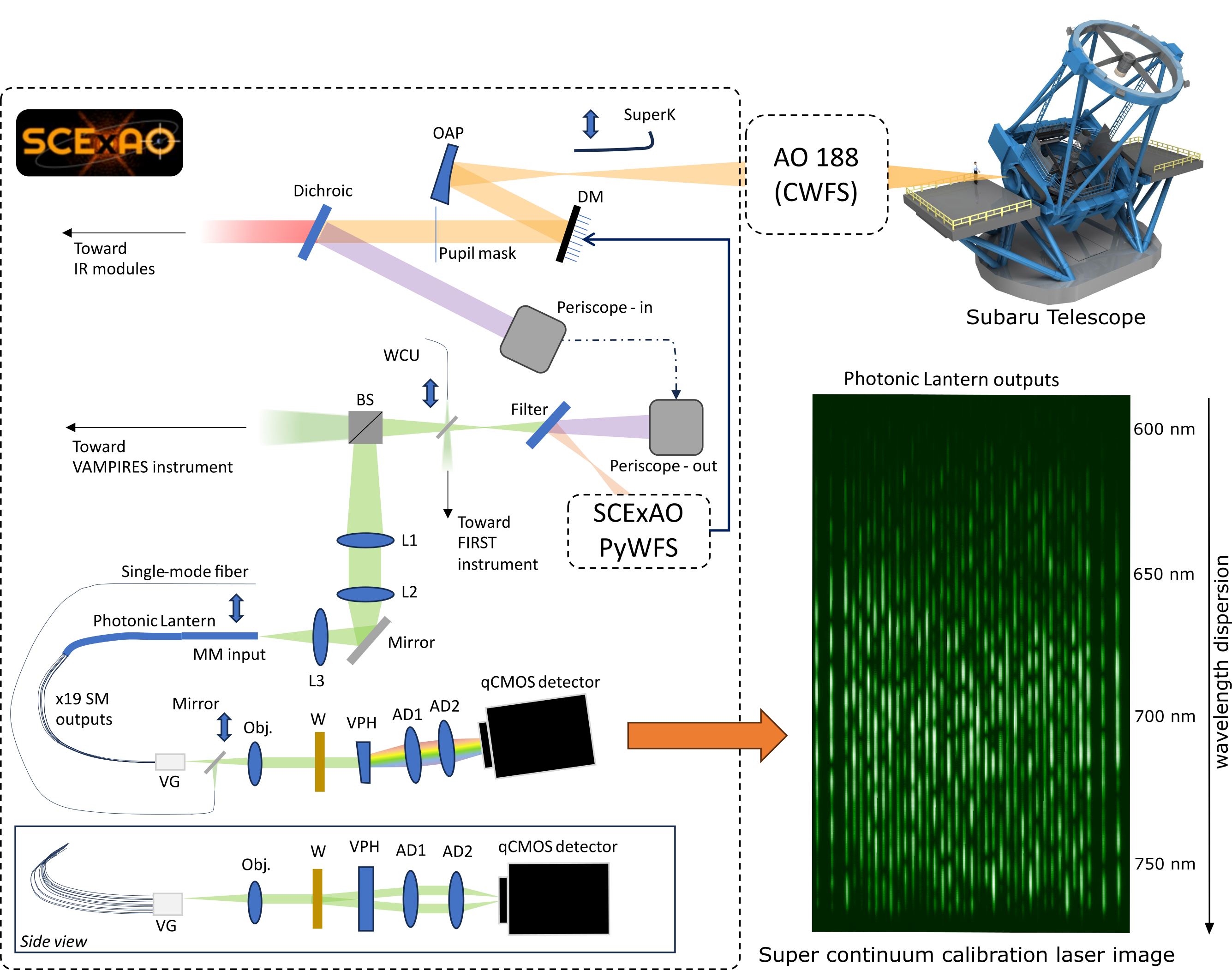}
        \caption{Integration of the PL at the Subaru Telescope. At the Nasmyth IR focus, light first enters AO188, where a first stage of AO correction is applied using a CWFS controlling a 188-actuator DM. The partially corrected beam then enters SCExAO, where an OAP collimates the beam. The beam is then reflected off a 2k BMC DM, then transmitted through a pupil mask mimicking the Subaru Telescope spider vanes. A dichroic splits light into a transmitted beam ($\lambda > 950$~nm) sent to the SCExAO IR modules, and a reflected beam ($\lambda < 950$~nm) sent to a periscope to transport the light toward a bench physically located above the near-infrared one. After, the output of the periscope, $\lambda > 800$~nm \rev{light} is reflected by a filter toward the pyramid wavefront sensor, performing the ExAO by controlling the 2k BMC DM. The transmitted beam is then picked up and sent toward the PL injection by a BS wheel. Upstream, a WCU can be inserted. The injection module is composed of a collimating lens (L1) and two focusing lenses (L2 and L3). Light can be injected in either the PL or a SMF thanks to a translation stage for wavefront quality calibration, and comparison with the PL. The MM input of the PL is split into 19 SMFs that are spliced to a linear VG. Each output is collimated by an Obj. The two polarizations of the beams are split by a W before being dispersed by a VPH grating. Finally, the dispersed outputs are imaged on a CMOS detector thanks to two achromatic doublets. A typical image acquired using the SCExAO SuperK calibration laser is displayed in the bottom right corner of the figure, where we can see the 38 traces coming from both polarized beam of the 19~SM outputs. The spectral dispersion here is horizontal. Similarly, the output of the SMF can feed the spectrograph by inserting a mirror, and generates two traces coming from both polarizations.}
        \label{fig:PL-on-scexao}
\end{figure*}{}

\begin{figure*}[h!]
        \centering
        \includegraphics[width=\linewidth]{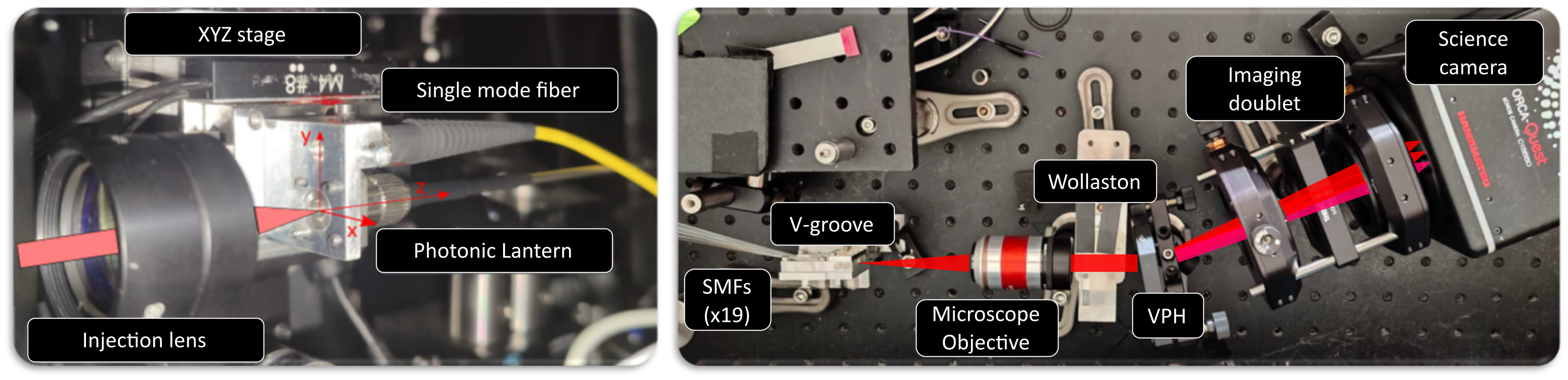}
        \caption{\corr{Opto-mechanical setup on SCExAO.} (Left) Picture of the injection module comprising the injection lens, the PL, and the SMF used for calibration. (Right) Picture of the R~3000 spectrograph. The SMF-fed spectrograph disperses the light thanks to a VPH grating and allows for polarization splitting of the signal thanks to a W.}
        \label{fig:PL-hardware}
\end{figure*}{}

\subsection{Instrument setup: Integration on SCExAO}
\label{subsec:Instrument_Setup_SCExAO}
The Subaru Coronagraphic Extreme Adaptive Optics~\citep[SCExAO - ][]{2015PASP..127..890J} instrument is located on the Nasmyth infrared (IR) platform at the 8.2-meter Subaru Telescope. SCExAO receives light from the Subaru Telescope AO facility instrument, AO188~\citep{minowa2010performance}. AO188 performs a first stage of AO correction using a curvature wavefront sensor (CWFS) controlling a 188-element deformable mirror (DM). SCExAO performs a second stage of AO correction, in order to reach the ExAO regime (with a Strehl ratio in the 0.8--0.9 range in the H band), enabling high-contrast imaging at high angular resolution and the detection of faint planets and brown dwarf companions \citep{kuzuhara2022,currie2023,tobin2024}. 
To achieve ExAO corrections, a pyramid wavefront sensor~\citep[PyWFS - ][]{Lozi_2019} operating in the $800$~nm to $950$~nm wavelength range drives a 2040-actuator MEMS DM. 

Figure~\ref{fig:PL-on-scexao} shows the optical configuration bringing the light from the telescope to the PL and the spectrograph. Light coming from AO188 is collimated by an off-axis parabola (OAP). A fiber carrying light from a Supercontinuum laser (hereafter named SuperK) can be inserted at the focal point for calibration or alignment purposes. The collimated light is then reflected by the 2040-actuator DM toward a pupil mask - mimicking the Subaru Telescope central obstruction and spiders for off-sky calibration. A dichroic mirror then splits the light, with near-infrared light ($\lambda>950$ nm) transmitted to the IR modules, and visible light ($\lambda<950$ nm) reflected  to a periscope. The periscope allows for the visible instruments to be mounted on an optical bench  above the bench hosting the IR modules. 

After exiting the periscope, visible light is first split by a dichroic filter: the reflected beam (from $800$~nm to $950$~nm) is sent to the pyramid WFS and the transmitted beam (<$800$~nm) is used by the visible modules of SCExAO. Before picking off the light for the PL, a R90/T10 filter can be inserted to either send the light to the FIRST spectro-interferometer~\citep{vievard2023singleaperture}, or send light from the wavelength calibration unit (WCU) to the PL. The WCU is a fiber connected to a neon lamp (AvaLight-CAL-Neon-
Mini), used for the PL spectrograph wavelength calibration. A beamsplitter (BS) wheel allows the beam to be reflected toward the PL or transmitted to the VAMPIRES dual-camera imager. The different slots of the BS wheel are currently occupied by a non-polarizing BS cube (NPBS), a polarizing BS cube, a $700$~nm short pass filter, a $750$~nm short pass filter, or an open slot. The following subsections present the injection module of the PL and the spectrograph.

\subsubsection{The injection module}
\label{sec:inj_module}


The purpose of this injection module is to efficiently inject light into the PL. The following experimental sections will show that, depending on the science application, there are several ways to inject light into the PL. The main parameter to be tuned is the input focal ratio. This focal ratio determines the field of view (FoV) or mode filling of the PL. The incident beam on the BS has a focal ratio of f/28.4.

Figure~\ref{fig:PL-on-scexao} presents the optical configuration to obtain a focal ratio of about 4. After the BS, the beam is collimated by a f$=$400\,mm doublet (named L1 in Fig.~\ref{fig:PL-on-scexao}). Two additional doublets of focal length 400\,mm (L2) and 50\,mm (L3) focus the beam onto the PL MM fiber input (see Fig.~\ref{fig:PL-hardware}~Left). A folding mirror located between L2 and L3 folds the beam for compactness and alignment requirements. Doublets are achromatic, making them well suited for injecting into the MM end of the PL. As is shown in Fig.~\ref{fig:PL-hardware}~(Left), the PL is mounted on a bracket composed of two~ports. In the second port, an SMF is installed for comparison between the two devices. \rev{The bracket is mounted on two remotely controlled translation stages (Zaber Technologies) for motion in the plane perpendicular to the optical axis (X and Y axes in Fig.~\ref{fig:PL-hardware}~(Left)), and on a conex stage for focus motion (Z axis in Fig.~\ref{fig:PL-hardware}~(Left)). They allow the PL input to be scanned across the focal plane of the injection lens, in order to optimize the light coupling into the PL.}

\subsubsection{The spectrograph}
The spectrograph is shown in Fig.~\ref{fig:PL-hardware} (Right). The 19 SM fibers of the PL output are spliced with SM600 type SMFs. The SMFs (hereafter referred to as PL outputs) are rearranged in a linear V-groove (VG) that aligns the fibers side by side in a pseudo-slit feeding the spectrograph. 
The distance between consecutive PL outputs is $127$~\textmu m for a total length of $2.286$\,mm. 

A 2$\times$ apochromatic microscope objective (Obj) with a 0.1 numerical aperture collimates the beams coming from the PL outputs. The collimated beams are then dispersed using a volume phase holographic (VPH) grating (manufactured by Wasatch Photonics). The transmission consists of two AR-coated glass covers protecting a layer of dichromated gelatin into which a $600$\,lines.mm$^{-1}$ interference pattern is written using coherent laser light. The VPH exhibits low scatter and low wavefront distortion. \rev{The manufacturer specifies that its dispersion efficiency, characterized between $600$ and $750$\,nm, is polarization-independent and ranges from $80$\% to $60$\%.} 
The grating is followed by a system of two imaging achromatic doublets with focal lengths of $150$\,mm (AD1) and $80$\,mm (AD2). 

Spectra are imaged on a high-frame-rate, low-readout-noise camera (Hamamatsu ORCA-Quest qCMOS). It contains $4096\times2304$ square pixels of $4.6$\,$\mu$m, a gain of about 0.11~$e^-$/ADU, and its quantum efficiency ranges from $78~\%$ to $58~\%$ between $600$ and $750$~nm. Finally, a Wollaston prism (W) is placed upstream of the grating to separate the spectra associated with two orthogonal linear polarizations on the camera. 

The resulting image is composed of 38~spectra coming from both polarized beams of the 19~SM outputs of the PL, as is shown in Fig.~\ref{fig:PL-on-scexao}. The spectra can then be combined (co-added) to reconstruct the source bulk spectrum.

\section{In-lab characterization}
\label{PL-inlab}

\rev{This section is dedicated to the characterization of our 19-port PL. After presenting some properties of the PL device itself, we present an extended study of ways to couple light into the PL and the associated device performance under various wavefront conditions.}

\subsection{Photonic lantern properties}

\begin{figure}[t]
\centering
	\includegraphics[width=\linewidth]{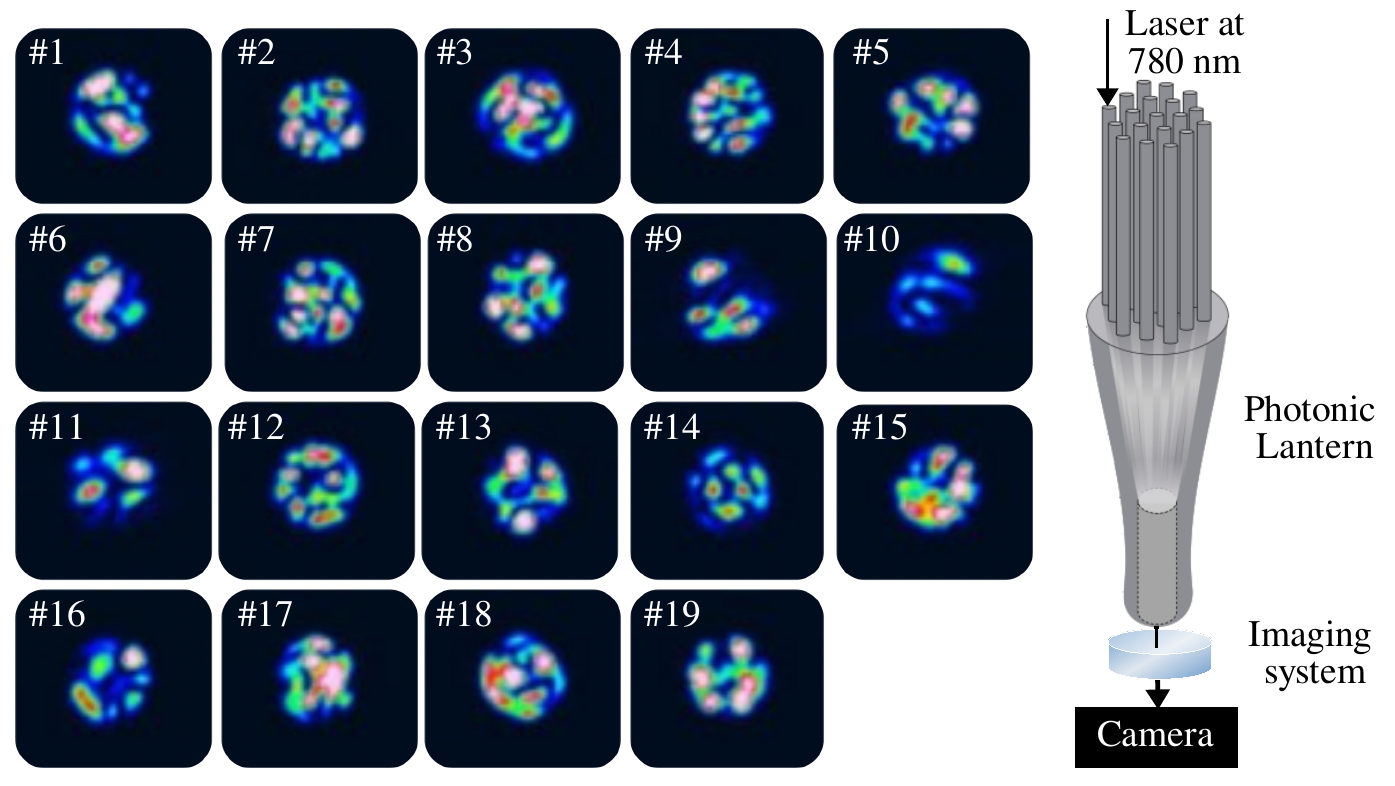}\\
        \includegraphics[width=\linewidth]{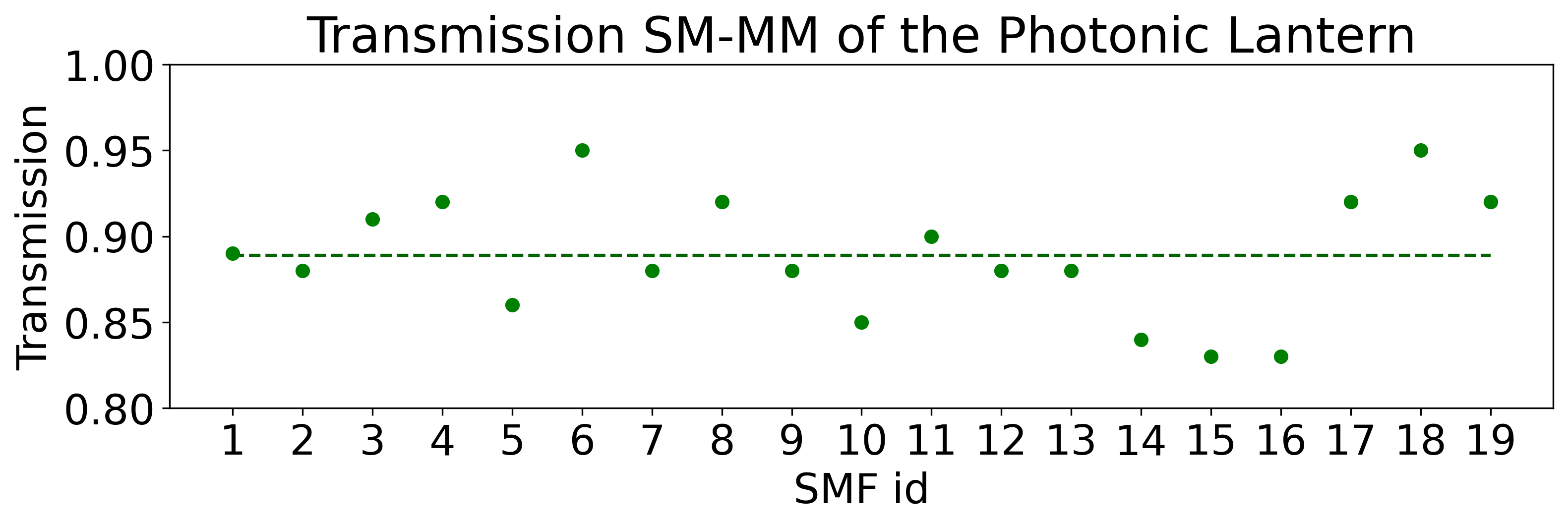}\\
	\caption{\corr{PL lab characterization.} Top panel: Distribution of the intensity of the light at the PL input after retro-injecting a $780~$nm laser in each of the PL SM outputs. Lower panel : Transmission from the SM end to the MM end of the PL, for each of the 19 SMFs.}
	\label{fig:PL_transmission}
\end{figure}

The images and transmission measurements shown in Fig.~\ref{fig:PL_transmission} were recorded using a laser diode tuned at 780nm (Thorlabs LPS-785-FC) with a measured standard deviation power stability of $0.3\%$ (over 2 hours) during the measurements. The transmissions values recorded in Fig.~\ref{fig:PL_transmission} were obtained by using a true cut-back method commonly used in optical fiber components. To perform the cut-back loss measurements, the output fiber of the laser was spliced to an SM PL pigtail and the output power at the PL MM end was recorded in a large area photo detector (Thorlabs S120C) by direct butt-coupling. Once the power was recorded, the SM PL pigtail was cleaved after the initial laser-SM PL splice and the power at the laser fiber end was recorded. The value of the transmission then is direct representation of the true transmission losses of the whole device by comparing those two measurements directly. This method is conveniently unaffected by either input coupling efficiency from the laser to the SM PL, or by the splice losses of the measurement assembly. It is important to note that the presented values are raw values of this method and do not include the possible Fresnel reflections at the MM PL end during the measurements – this could be as high as $4\%$ on the silica-air interface. The recorded laser stability during the measurements and the significant figures of the large area photo detector apparatus gave us an error of $0.5\%$ on the recorded transmissions.

\rev{The images shown in Figure~\ref{fig:PL_transmission} represent the 19 different orthogonal superpositions of the LP (modes guided by the lantern MM end) that map to each of the SM PL inputs independently. These are unique and orthogonal solutions for a fabricated PL that depends on wavelength.}

The average measured transmission is $88.9\%$\rev{, with values spanning between $83\%$ and $95\%$}. This is in accordance with the expected $>90\%$ transmission expected in a quasi-adiabatic PL transition between MM and SM ends~\citep{birks2015photonic}.

{}


\subsection{Coupling light into the photonic lantern}



\subsubsection{Injection optimization using coupling maps}

The injection optimization into the PL was performed \rev{by scanning the focal plane in the (X,Y) directions by using the translation stages mentioned in Sect.~\ref{sec:inj_module}}. A scene window was defined in units of translation stage steps with two parameters: the total width of the \rev{scanned} window and its sampling. The stages then positioned the PL at each sampled (X,Y) coordinate, and for each location, an image of the PL outputs was acquired with the spectrograph. After the scan was complete, coupling maps were computed, corresponding to the total flux acquired for each (X,Y) position. \rev{The source used for this test was the SuperK.}

\begin{figure}[t]
\centering
        \includegraphics[width=\linewidth]{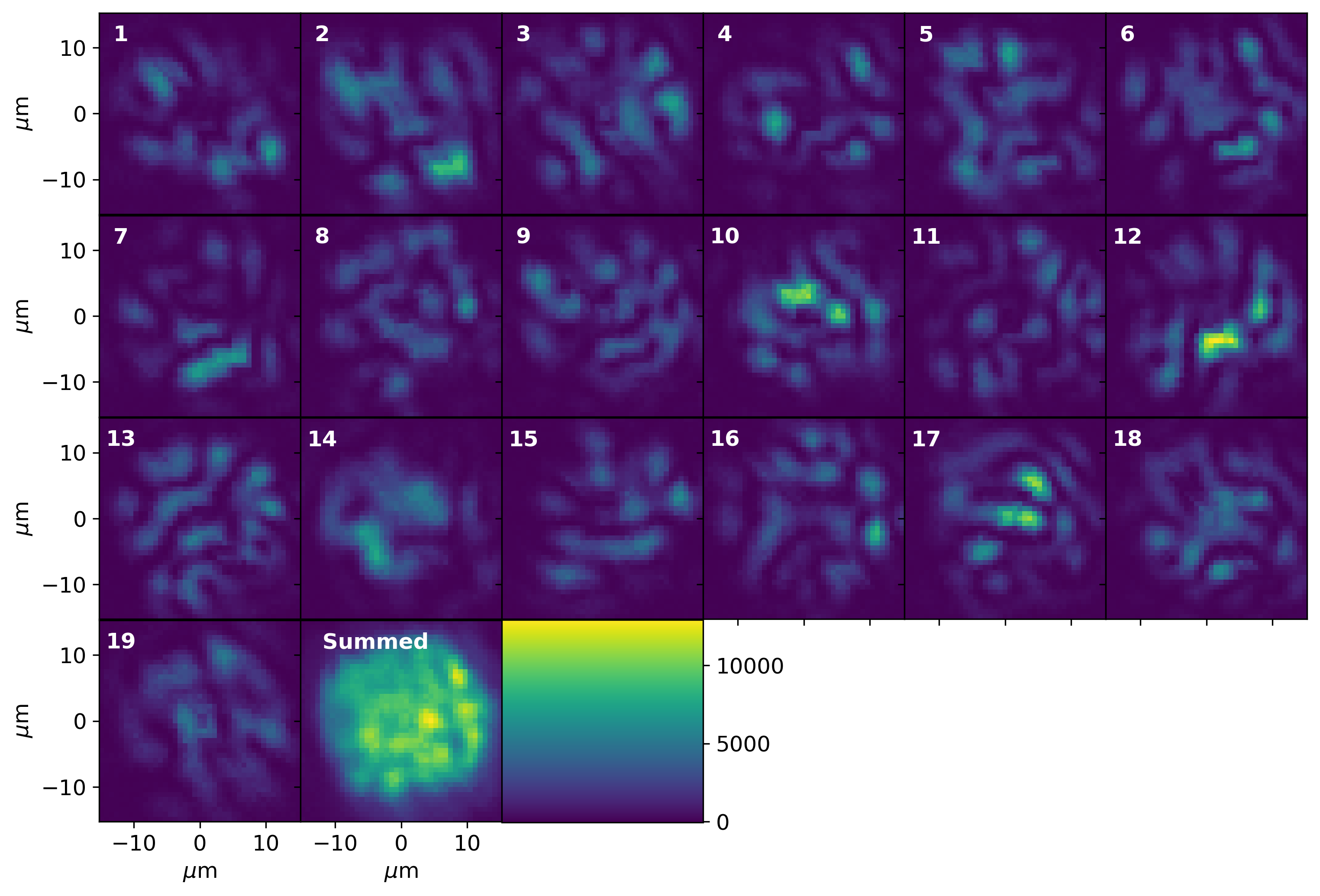}
        \caption{Coupling maps computed for all 19~outputs, acquired for a focal ratios of 4 and at a wavelength of 765.5~nm. The coupling map for each output presents some fine high-order structures, similarly to the light distribution when imaging retro-injected light into the SM outputs. The sum of the coupling maps for each wavelength provides a smoother pattern.}
        \label{fig:PL_coupl}
\end{figure}

This process was performed simultaneously for each output and each wavelength, providing $n_o \times n_\lambda$ coupling maps, with $n_o$ the number of outputs (here, 19) and $n_\lambda$ the number of spectral channels (here, 1896 spanning from $593.5$~nm to $785.5$~nm). In addition, we compiled the total coupling map, defined as the sum of all 19~output maps, capturing the global behavior of the PL. Figure~\ref{fig:PL_coupl} presents the resulting coupling maps for the 19 outputs, at a single wavelength (765.5~nm). \rev{This wavelength was chosen because it presented the best S/N.} The injection setup was fixed with a focal ratio of 4.

{}

All coupling maps exhibit high-order structures, similarly to the light distribution imaged at the PL input when retro-injecting in each SM output (see Fig.~\ref{fig:PL_transmission}). The summed map is smoother, but still granulated. The darkest “spot” of the total map is $40\%$ fainter than the brightest spot. The size of the map envelope is about 26~$\mu m$, which is comparable to the beam size of the fundamental mode (about $22.5~\mu m$) convolved with the PSF (about $3~\mu m$ at 765~nm with a focal ratio of 4). 

\begin{figure}[t]
\centering
	\includegraphics[width=\linewidth]{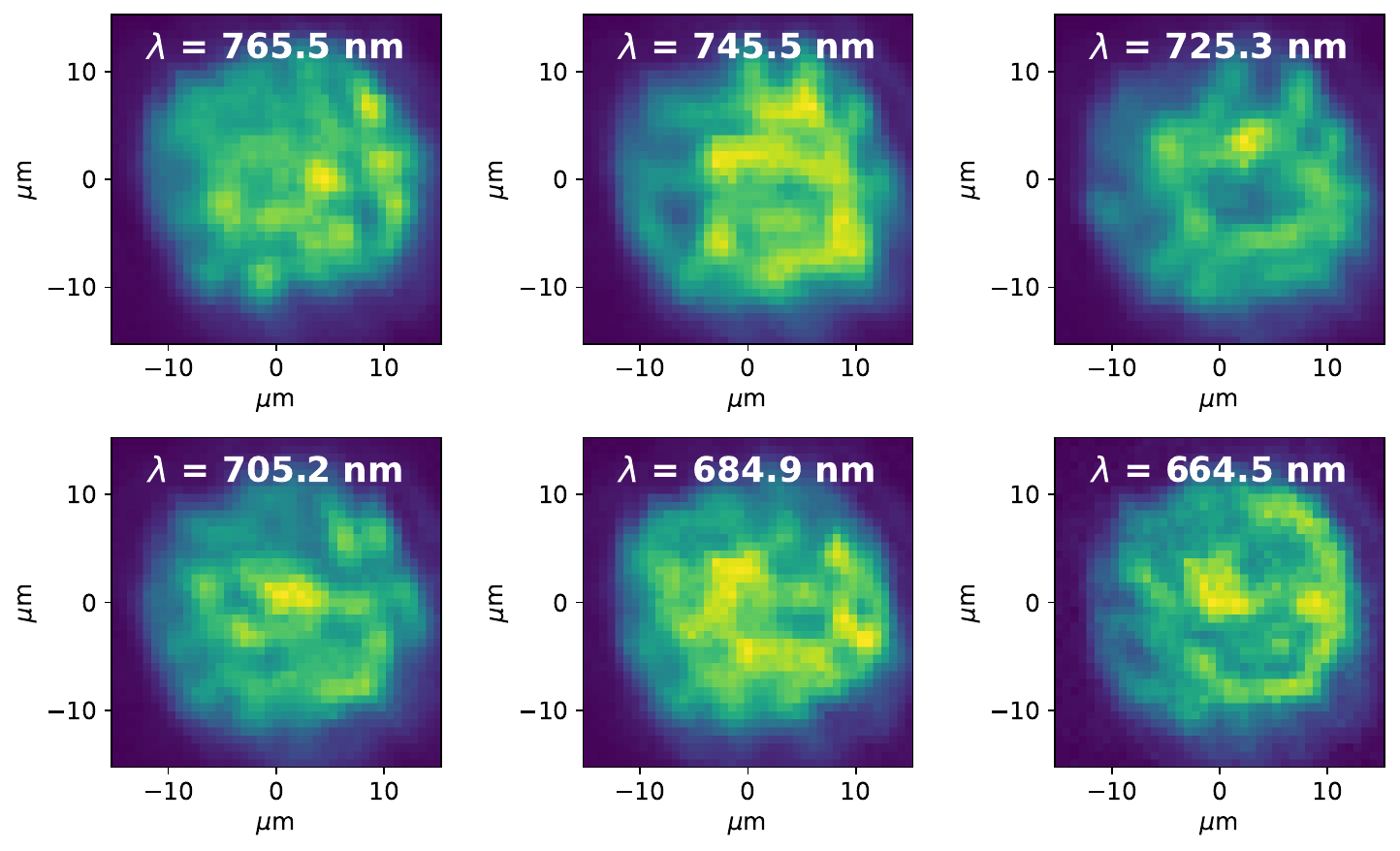}\\ 
	\caption{Total coupling maps obtained for different wavelengths at a fixed focal ratio of 4. \rev{False colors are normalized the same way as in Figure~\ref{fig:PL_coupl}}.}
	\label{fig:PL_coupling_wl}
\end{figure}

Once the coupling map was  acquired, a 2D Gaussian function was fit to the coupling map to find the coordinates of its center, corresponding to the optical axis, or the FoV center. We then drove the PL stages to these coordinates. Alternatively, one can choose to position the PL at another location; for example, to maximize the flux injected into the PL by choosing the coordinates corresponding to the brightest spot of the summed coupling map. The summed coupling map is granulated, which means that the throughput is not homogeneous across the field.  Figure~\ref{fig:PL_coupling_wl} presents the PL coupling maps summed over the 19 outputs, for six different wavelengths. The comparison of these maps shows that the granulation pattern changes with wavelength.

{}

\begin{figure}[t]
\centering
        \includegraphics[width=\linewidth]{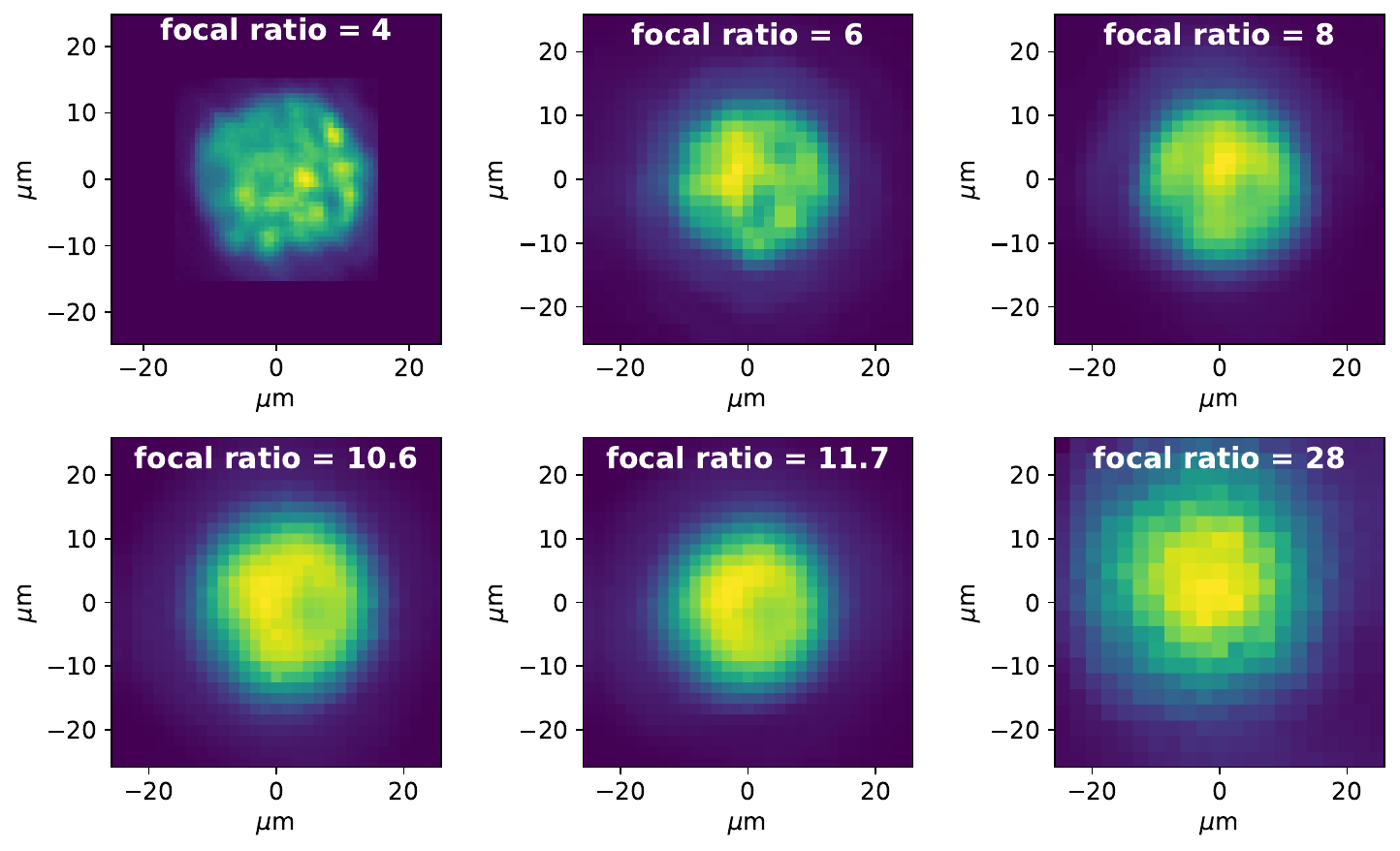}
        \caption{ Coupling maps acquired for various focal ratios at a fixed wavelength of 765.5~nm..}
        \label{fig:PL_coupling_fratio}
\end{figure}

Because the size of the PSF is small compared to the \rev{fundamental MM PL input mode} (by a factor of about 7.5 \rev{at 765~nm}), the number of modes of the input wavefront is too large compared to the output 19~modes. An over-sampled input mode field at the input of a few-mode fiber (i.e., much smaller input PSFs diameters than the guided mode field diameters of the few-mode fiber) will result in a spatially and wavelength coupling efficiency mismatch and dependence on the input coupling conditions, leading to the overall granulation observed in the summed coupling map. Figure~\ref{fig:PL_coupling_fratio} shows the total coupling map granulation for different focal ratios. The larger the focal ratio, the less granulated the coupling map. This is explained by the fact that a larger PSF (or larger focal ratio) corresponds to fewer modes sampled at the PL injection.

{}

\setcounter{figure}{10-1}
\begin{figure*}[b]\centering
\begin{tabular}{cc}
(a) Focal ratio = 4 & (b) Focal ratio = 6 \\
        \includegraphics[width=0.47\linewidth]{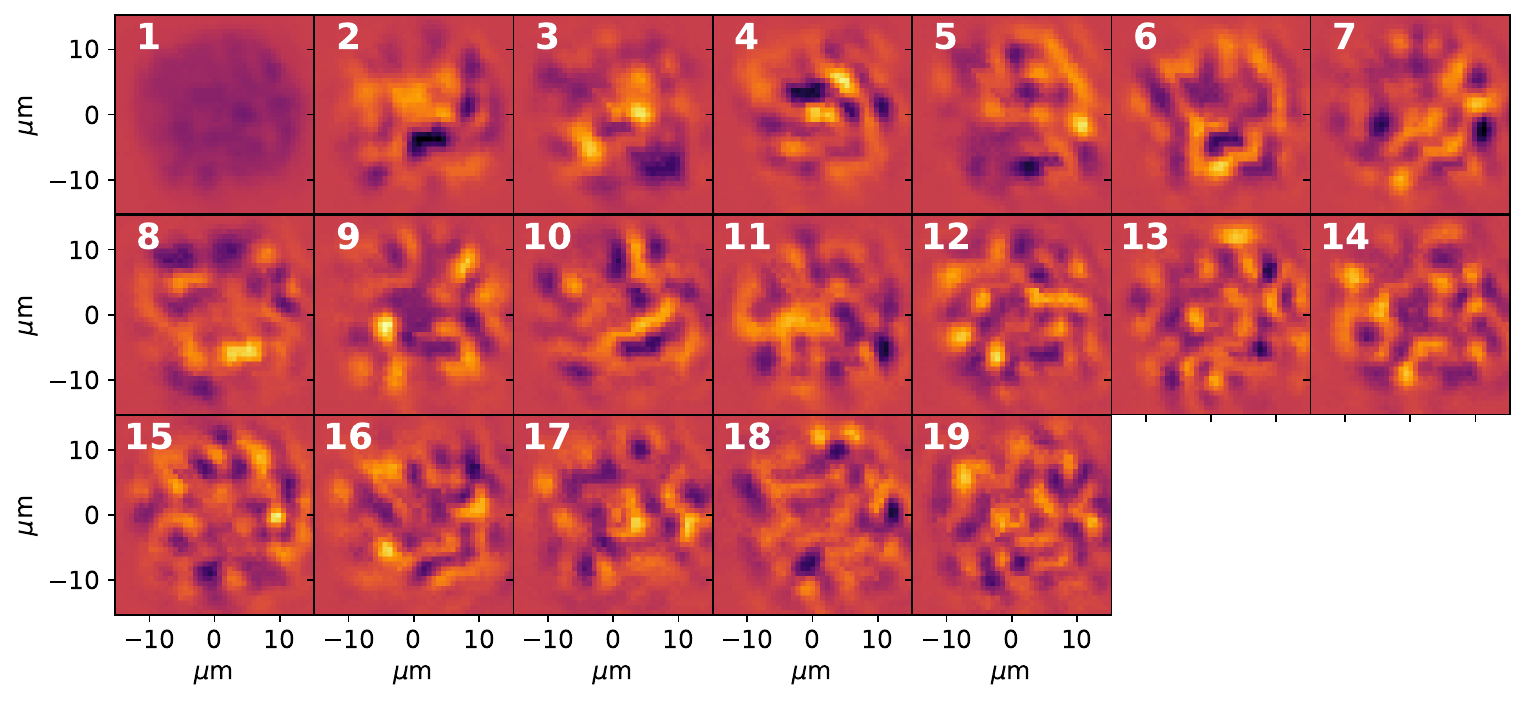} &
    \includegraphics[width=0.47\linewidth]{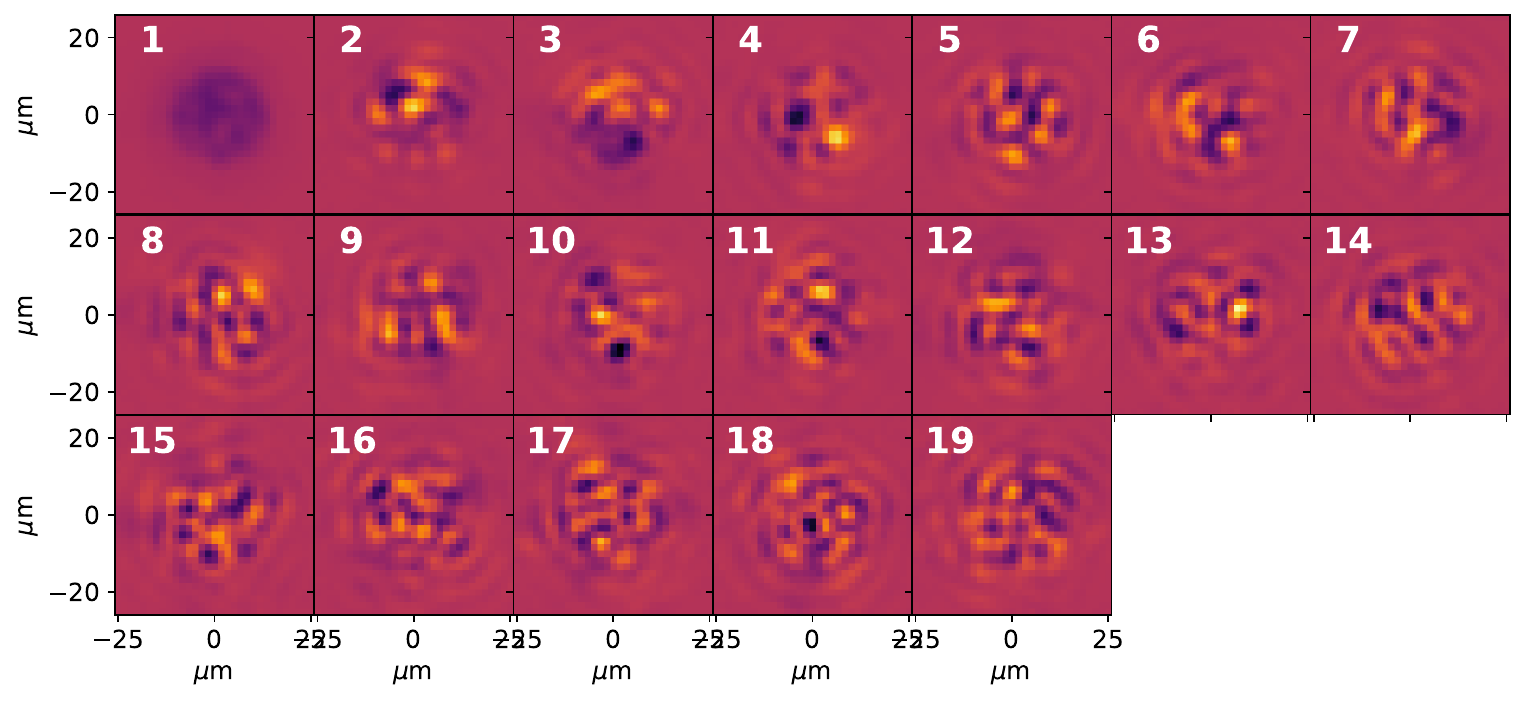}\\
(c) Focal ratio = 8 & (d) Focal ratio = 10.6 \\
    \includegraphics[width=0.47\linewidth]{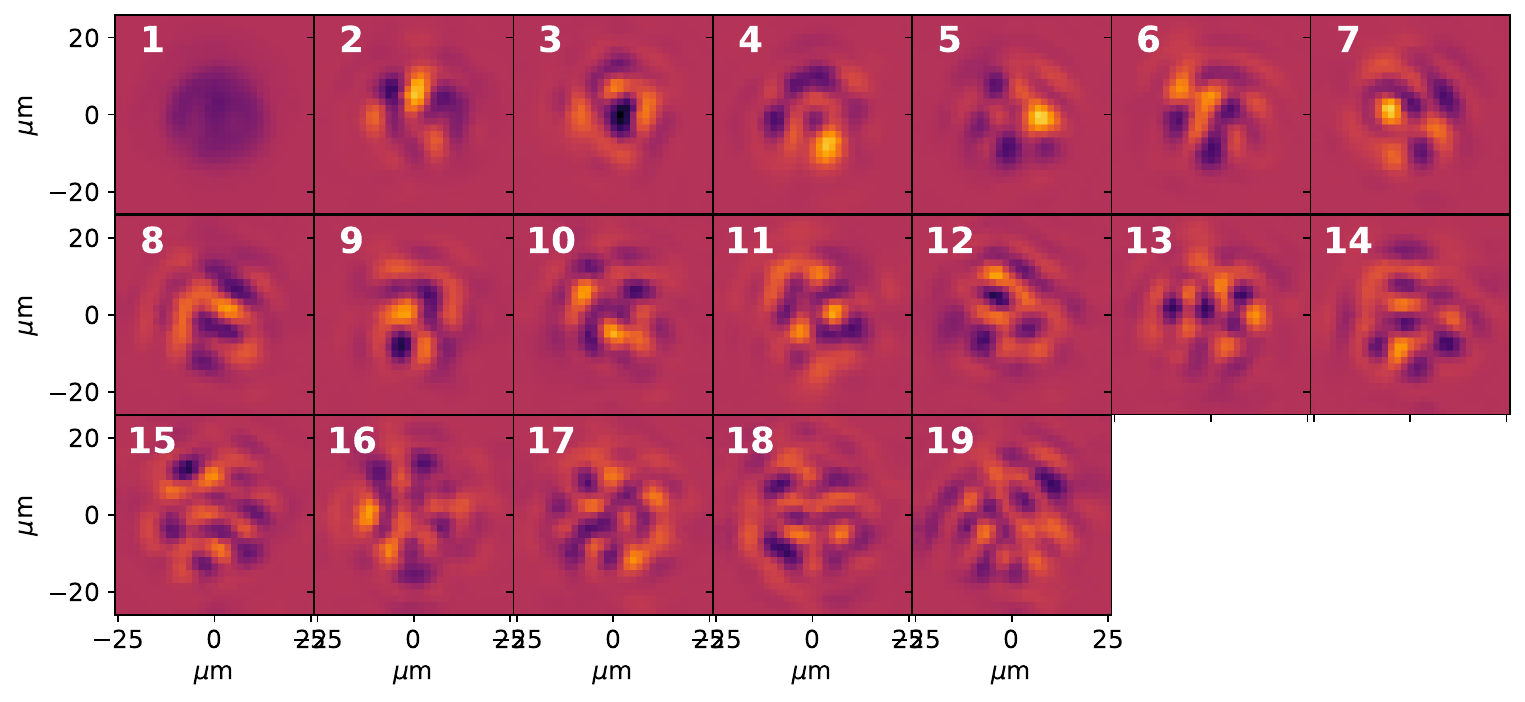} &
        \includegraphics[width=0.47\linewidth]{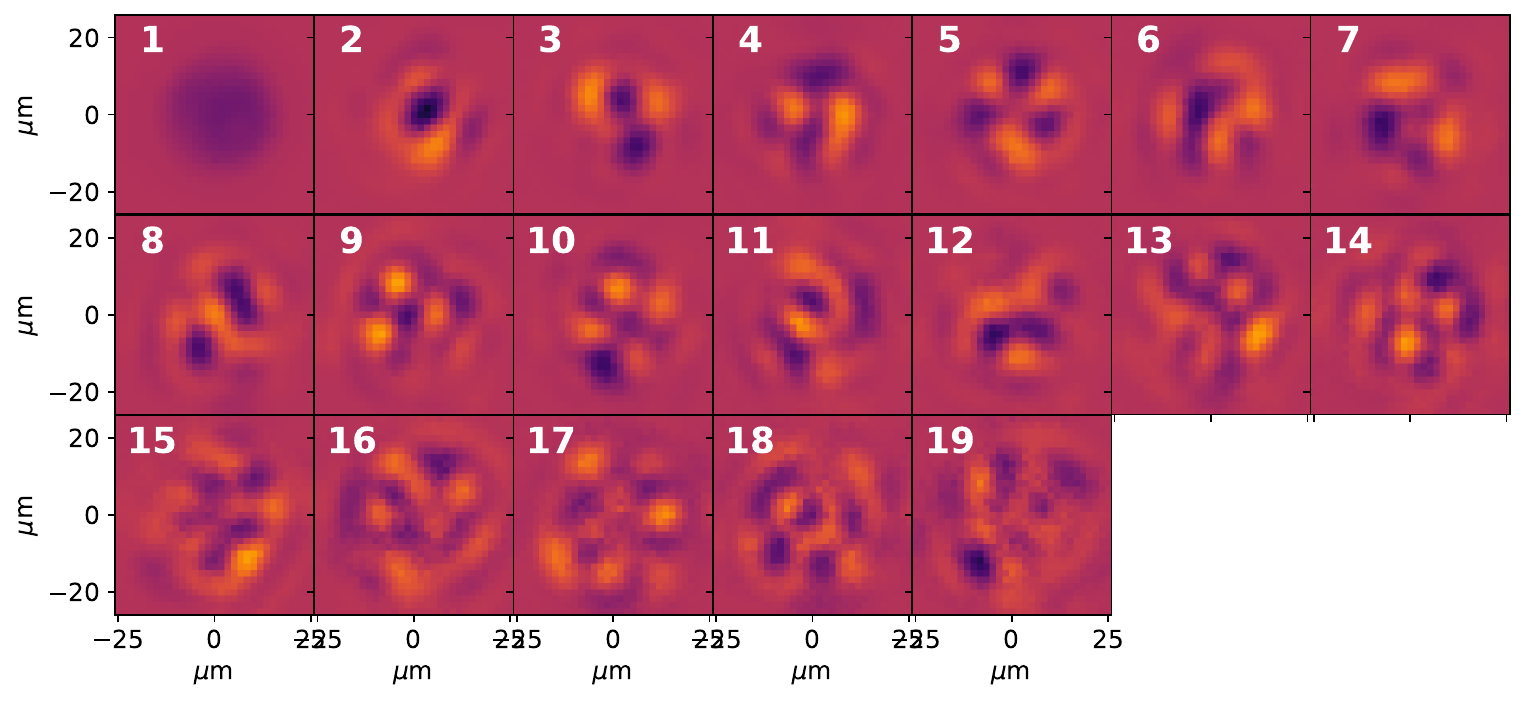}\\
(e) Focal ratio = 11.7 & (f) Focal ratio = 28 \\
    \includegraphics[width=0.47\linewidth]{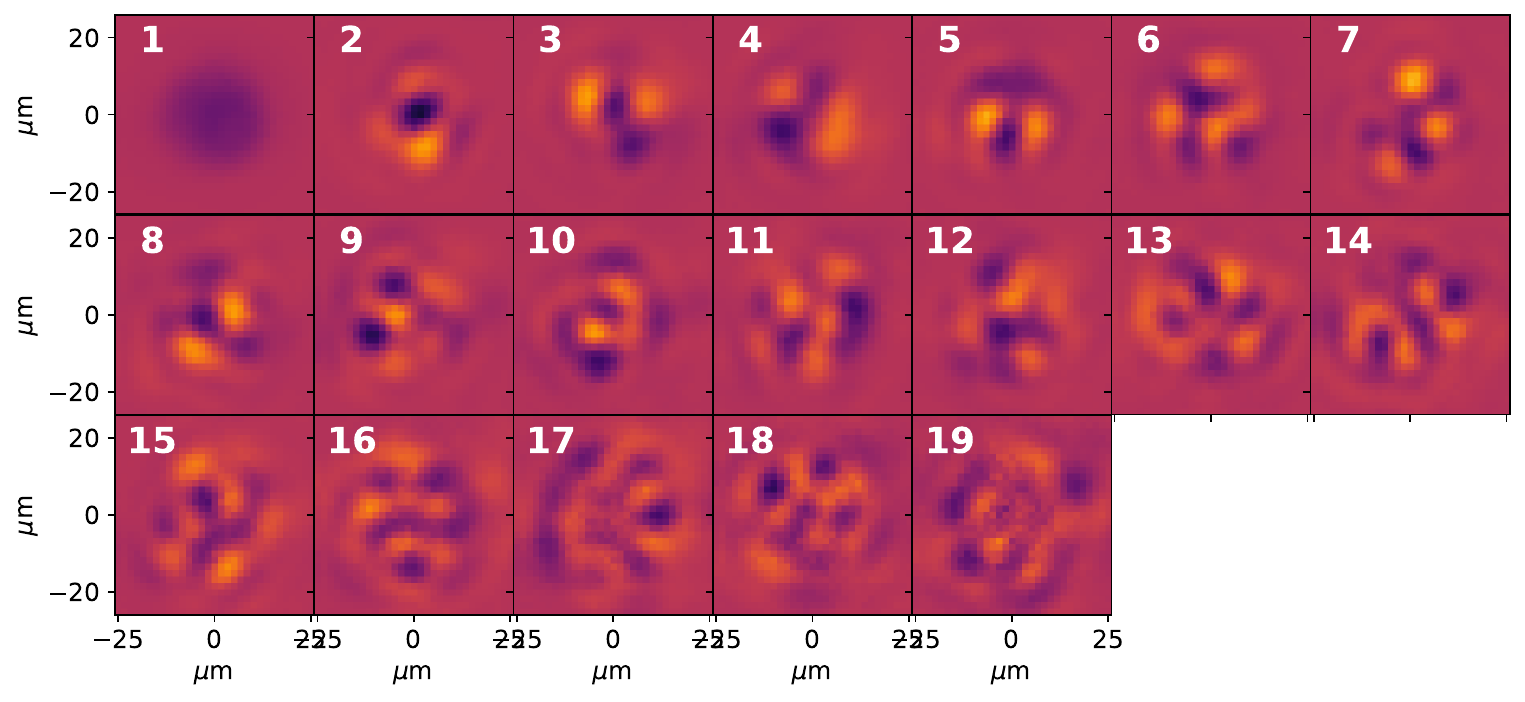} &
        \includegraphics[width=0.47\linewidth]{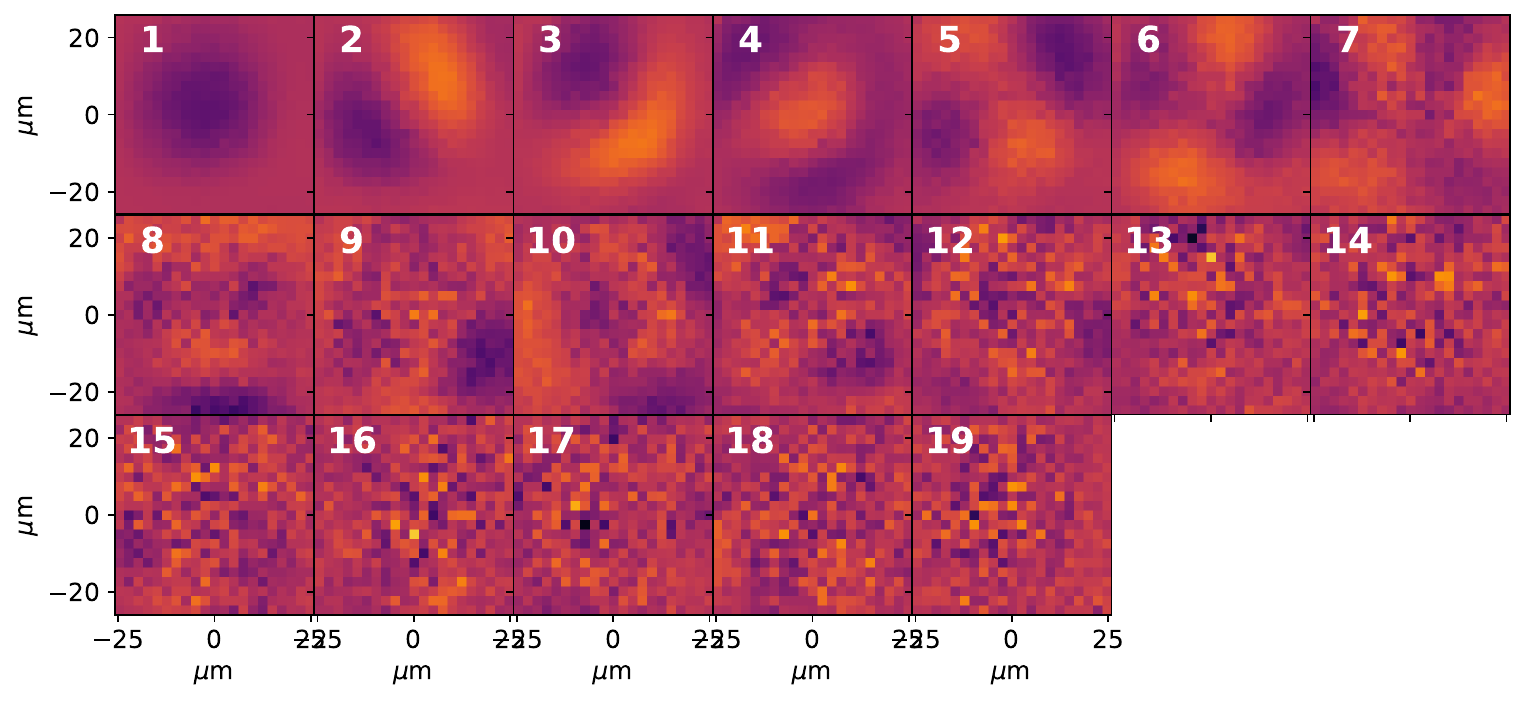}\\
\end{tabular}
\caption{Modes seen by the PL from the coupling maps at one wavelength ($\lambda = 765.5$~nm) for focal ratios ranging from 4 to 28.}\label{fig:PL_modes2}
\end{figure*}

\setcounter{figure}{9-1}
\begin{figure}[t]
\centering
        \includegraphics[width=\linewidth]{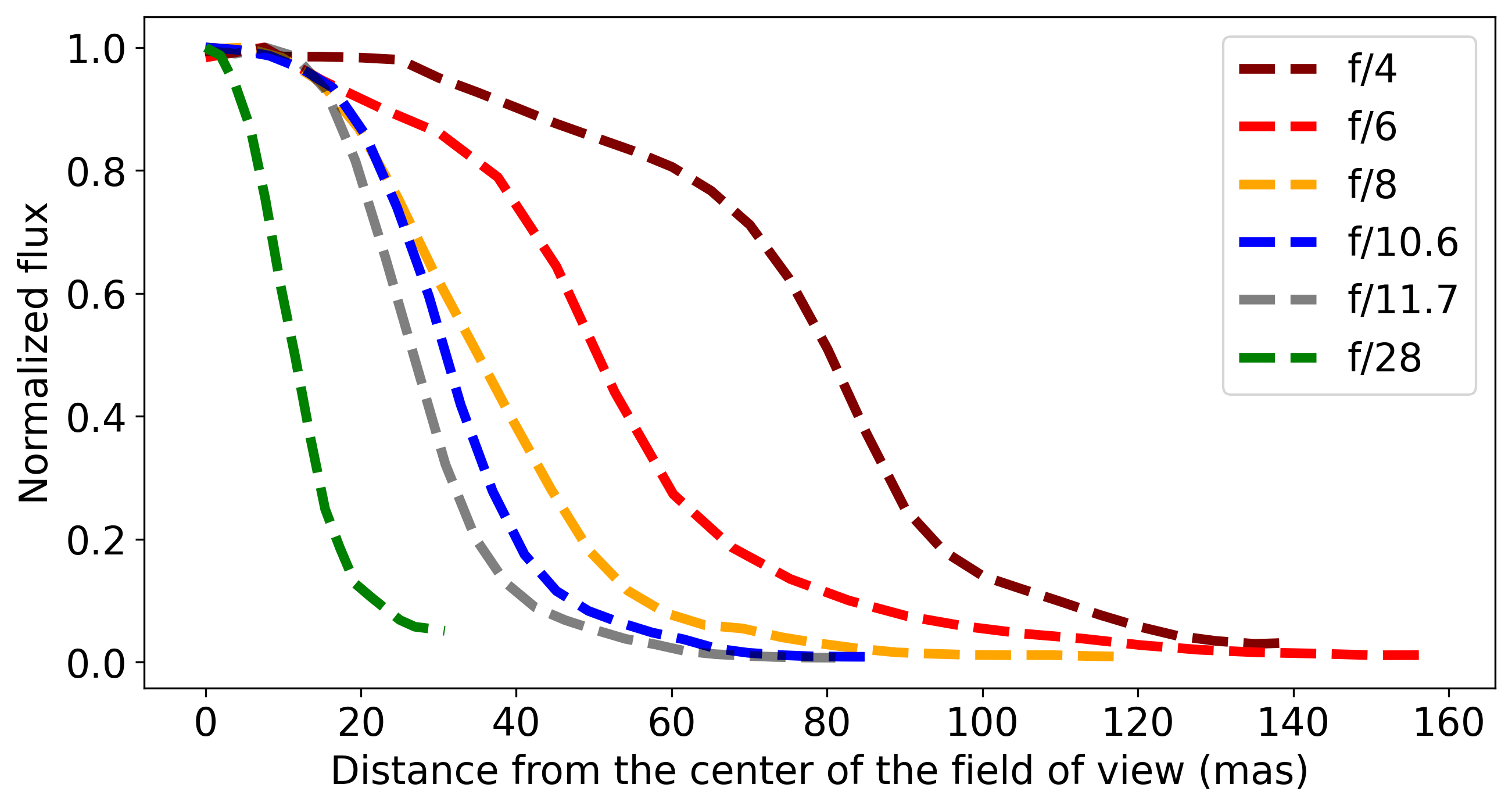}
        \caption{Field of view of the PL for various f-ratios \rev{computed from the radial average of coupling maps}.}
        \label{fig:PL_fov}
\end{figure}{}

\setcounter{figure}{11-1}

\subsubsection{Field of view}

The coupling maps are not only a proxy to optimize the injection into the PL; they can also inform on the FoV when converting the measured coupling maps from linear motion coordinates to sky angular coordinates. We subsequently computed the radial average of the resulting map to estimate the angular FoV. The resulting curves, shown in Fig.~\ref{fig:PL_fov}, have been normalized and show that the FoV increases when the focal ratio decreases. This can be taken into account when choosing the science application for the PL.

\subsubsection{Parametric analysis of injected modes via singular value decomposition}

We applied a singular value decomposition (SVD) to the coupling maps provided by the 19~outputs of the PL, for each tested focal ratio. To do so, each of the 19 coupling maps of dimension [$n_x$,$n_y$] was converted into a vector of dimension $n_x\times n_y$. We then constructed a matrix, $A$, with dimensions [$19,n_x\times n_y$]. The SVD was applied on the resulting matrix, $A$, and provides the SVD modes (that we name injection modes) and singular values associated with each mode. 

The modes are shown in Fig.~\ref{fig:PL_modes2}, noted from 1 to 19, for focal ratios ranging from 4 to 28\rev{, and for one wavelength ($\lambda$~=~765.5 nm)}. Apart from the first mode, which is an average of the global envelope, we see that the modes have more high-frequency structures when the focal ratio decreases (hence, the PSF size decreases). For focal ratio values between 4 and 11.7, the lantern couples 19 input modes. For a focal ratio value of 28, there are only roughly seven modes injected, and modes 8 to 19 seem to be mostly noise -- consistent with the low singular values. \rev{Similar behavior is observed for the rest of the spectrum.}

\begin{figure}[h]
\centering
        \includegraphics[width=\linewidth]{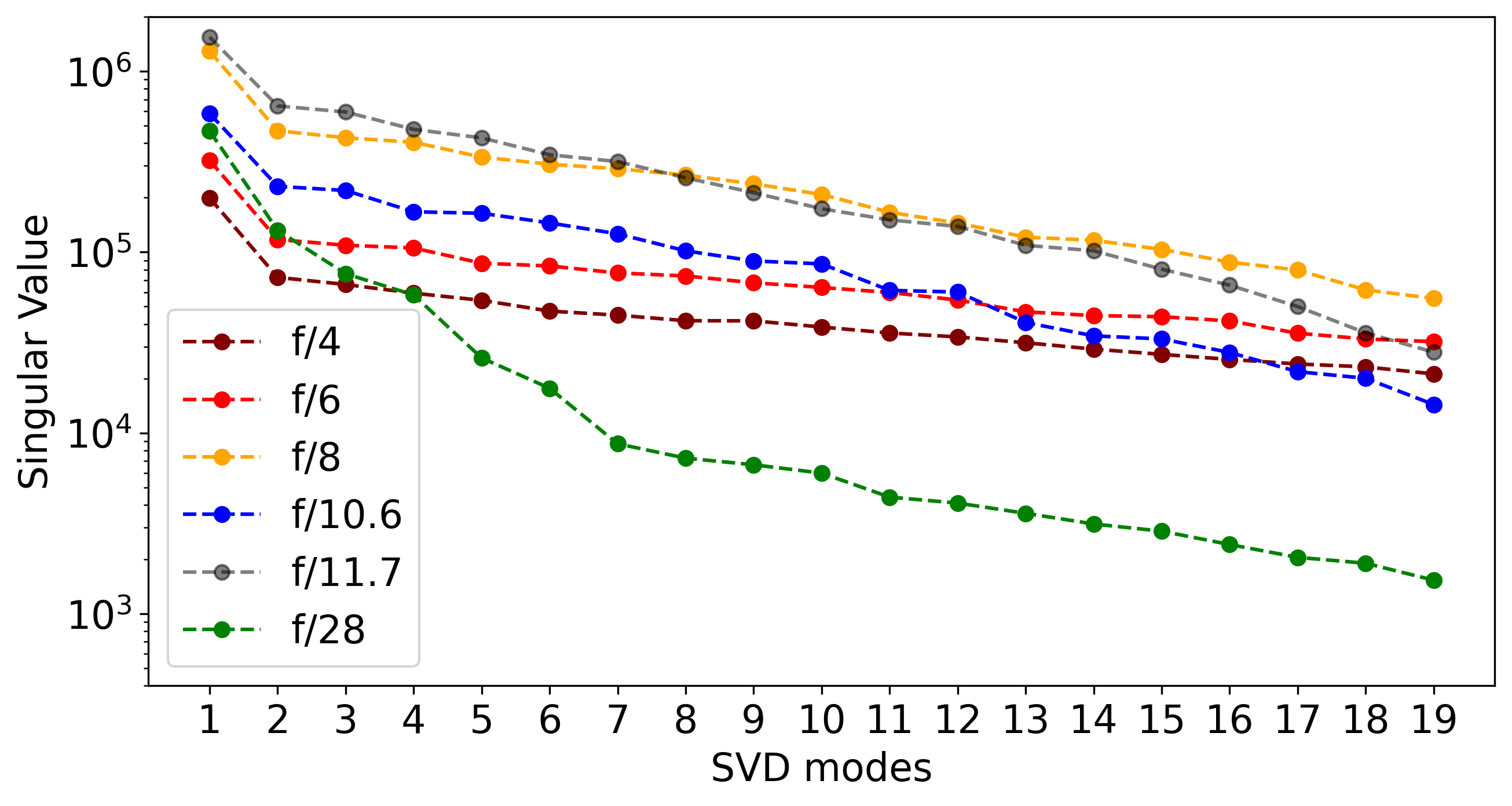}
        \caption{Singular values of the modes injected into the PL for various f-ratios.}
        \label{fig:PL_SV}
\end{figure}

\rev{Figure~\ref{fig:PL_SV} shows the singular values computed for each SVD mode. Interestingly, the singular values for most of the SVD modes in the f/8 and f/11.7 cases are higher than those in the other cases. This indicates that the modes generated in the f/8 and f/11.7 cases are more prominently represented in the data compared to the other cases. As a result, f/8 and f/11.7 might be better suited for the system, as they capture more of the data's variance and provide a more accurate representation.}

{}

\rev{We can also deduce} from the graph that the matrix conditioning number --- defined as the ratio of the largest singular value to the smallest singular value ---  increases with the focal ratio. This indicates that the stability of the inversion diminishes as we progressively introduce fewer “genuine” input modes (or as the f-ratio increases). Among these modes at high f-ratio, some are authentic modes observed by the PL, while others merely constitute noise, as is discussed below. \rev{This modal study is a stepping stone toward a linear representation of the PL, and will be used in the future to investigate image reconstruction using the PL.}

\subsection{Throughput characterization}
\subsubsection{Injection versus wavelength}
\label{sec-injvswl}
\begin{figure}[h]
\centering
        \includegraphics[width=\linewidth]{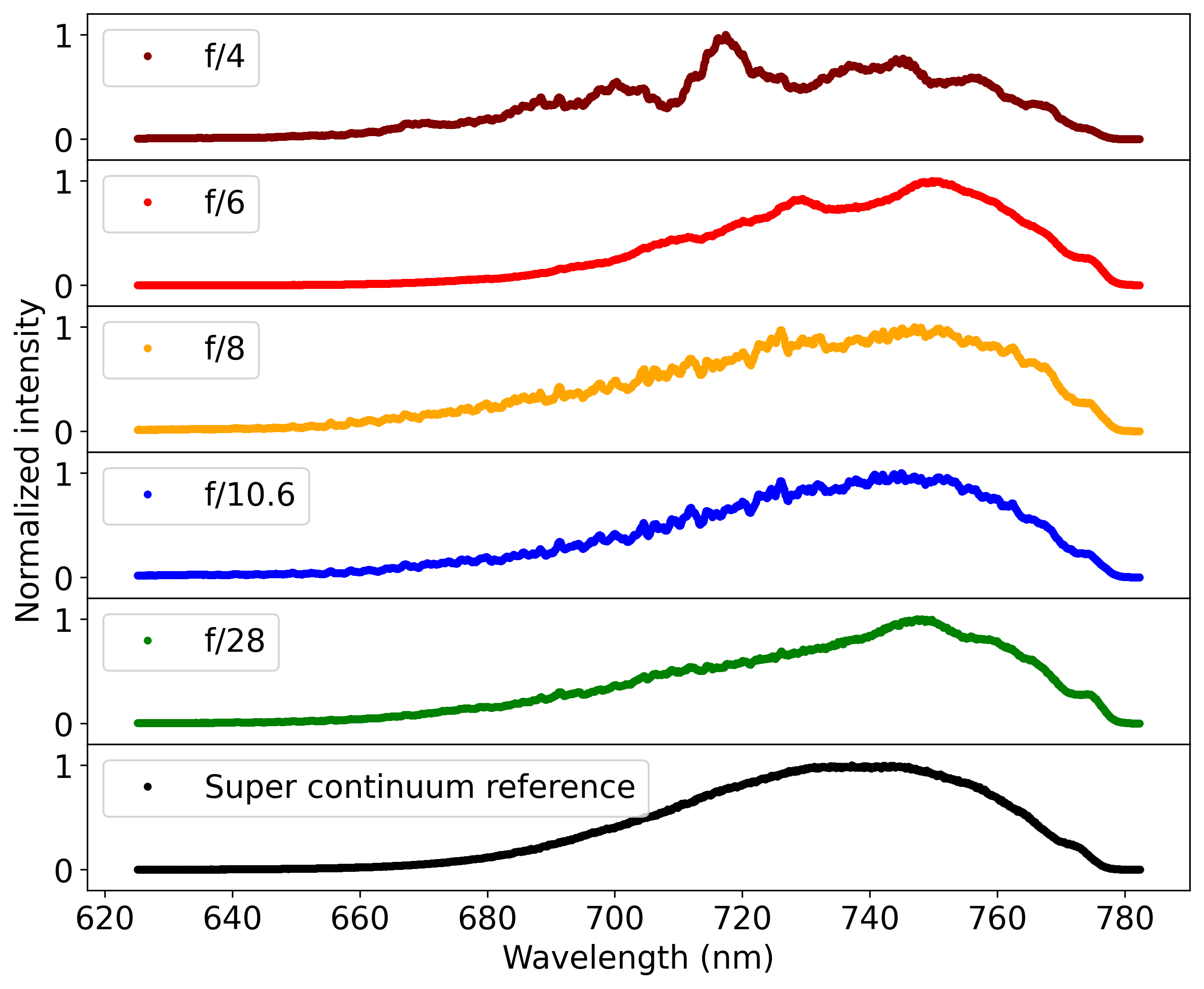}
        \caption{Super continuum spectrum reconstructions from PL data at various focal ratios. Each spectrum is normalized by its maximal value. The spectra can be compared to the reference on the bottom plot, obtained with an SMF.}
        \label{fig:PL_spectra}
\end{figure}

\rev{We now study the spectrum reconstruction of data acquired with the PL on the SCExAO super-continuum calibration source. We followed the previously mentioned injection optimization and placed the PL input at the center of the coupling map in each case.}
Depending on the application and noise regime, different spectrum extraction methods can be considered: from simply co-adding all the traces to performing a weighted signal extraction on each output, where only the bright parts of each trace are summed. In the photon-noise limited regime, a simple co-addition of the traces is optimal. In a read-out-noise-limited regime, minimizing the number of pixels leads to higher signal-to-noise (S/N), so a weighted signal extraction method should be considered. For the purpose of this paper, we shall not discuss the various spectral extraction methods, and have simply chosen to extract and co-add the 38~traces. \rev{This is especially suitable for objects characterization through spectroscopy, but also for radial velocity measurements.} We also acquired a reference spectrum using the SMF mounted on the same injection setup as the PL. We present the reconstructed data in Fig.~\ref{fig:PL_spectra}.

{}

\begin{figure*}[t]
\sidecaption
        \includegraphics[width=12cm]{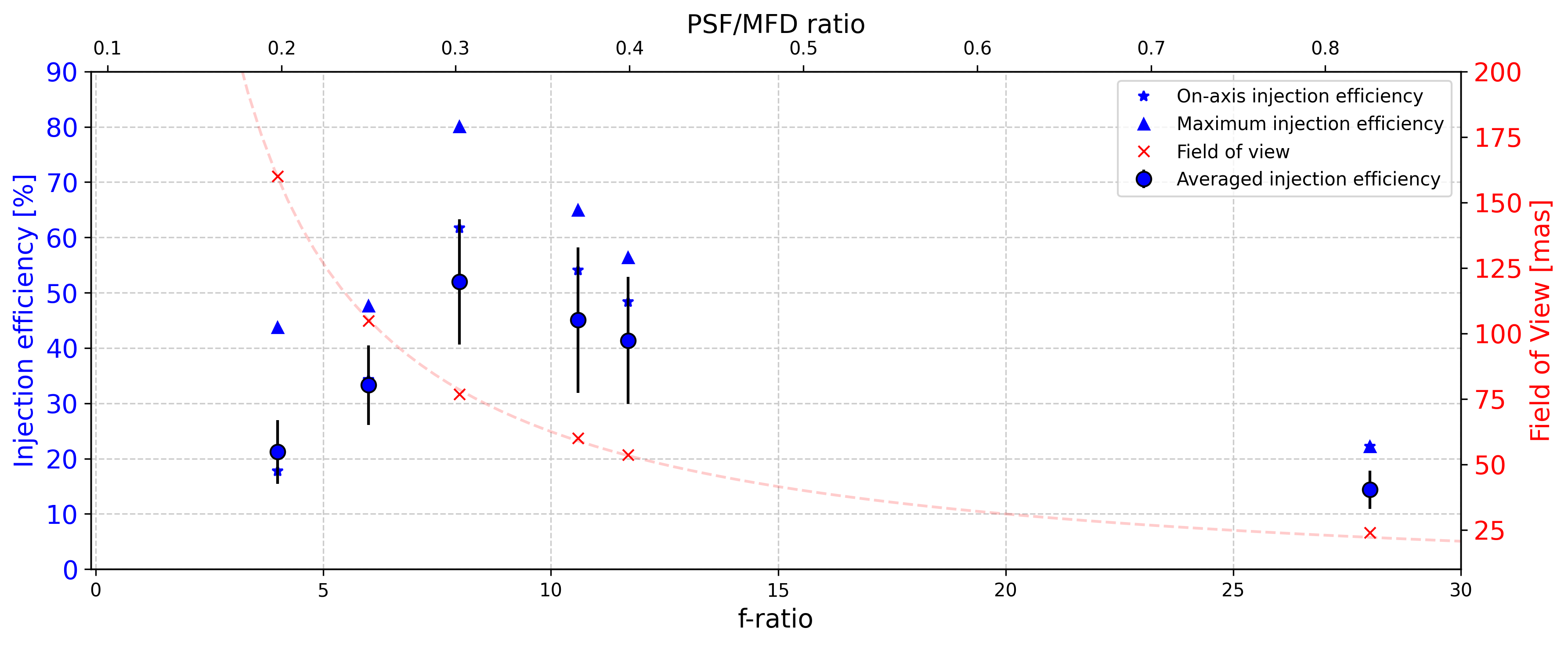}
        \caption{Variation of the injection efficiency measured at 642~nm as a function of the focal ratio (bottom horizontal axis), or as a function of the ratio between the PSF size and MFD of the PL (top horizonal axis). For each focal ratio experimentally tested, we represent the on-axis injection efficiency, the maximum efficiency, and the average over the whole scanned area (see the coupling maps). The FoV projected on-sky is also plotted with red crosses (right axis).}
        \label{fig:PL_coupling}
\end{figure*}{}

\begin{figure*}[h]
\sidecaption
        \includegraphics[width=12cm]{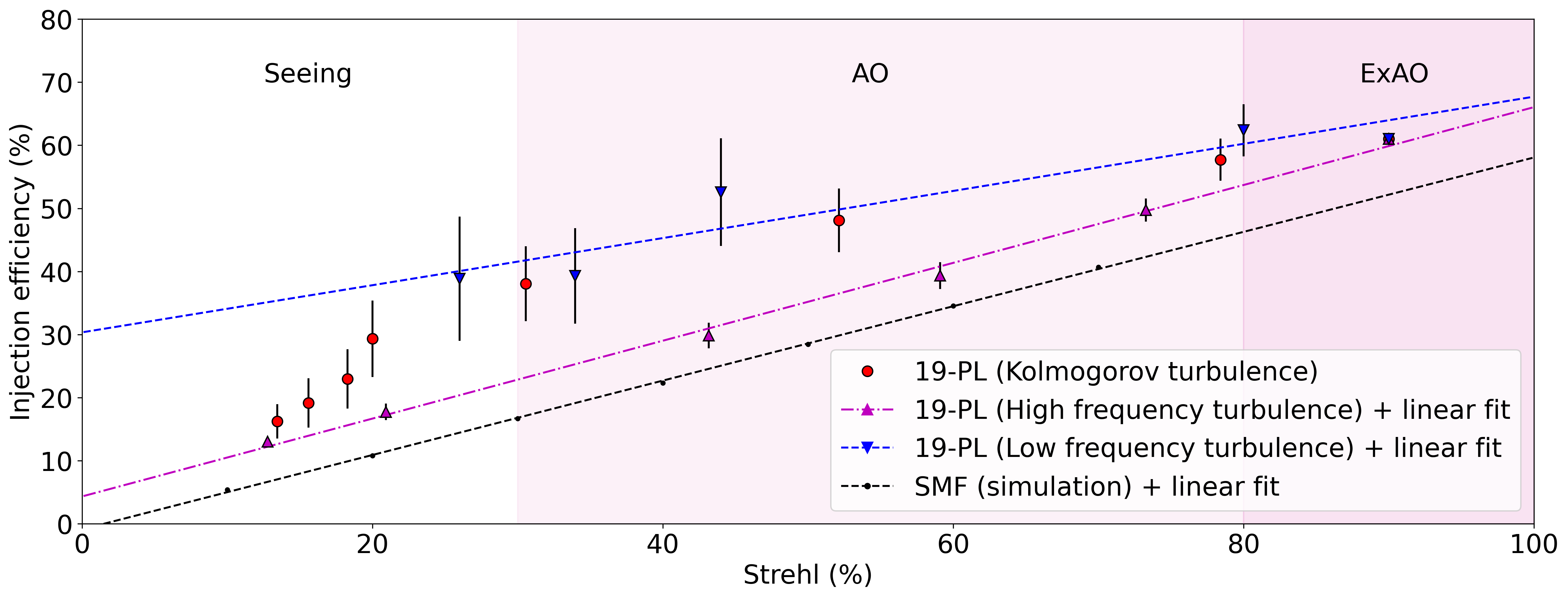}
        \caption{Relationship between the injection efficiency \revv{at 642~nm} and the Strehl ratio \revv{measured at 750~nm} for various atmospheric conditions. The turbulence is applied on the SCExAO DM and the flux is recorded at the 19-port PL output. The presented SMF results are simulations from \cite{lin2021design}\revv{, where the simulated turbulence screens were following the Kolmogorov power spectrum}.}
        \label{fig:PL_injeffstrehl}
\end{figure*}{}

Comparing the PL reconstructed spectra to the reference spectrum reveals some low-frequency modulation for focal ratios 4 and 6. These correspond to a loss of flux between the input beam and the output of the PL. In the case of low focal ratios, due to the small PSF size there is more energy in high-frequency spatial modes that are not supported by the 19 modes coupled from the MM input to the SM outputs, and hence cannot be injected leading to wavelength dependence losses in the overall spectra. 
This effect needs to be taken into account since it could make the continuum reconstruction challenging depending on the targeted science case. The spectra acquired for larger focal ratio do not exhibit these large modulation patterns: the spectral reconstruction is more accurate.

Some high-frequency fringes are also observed in all of the PL spectra, but are not present in the SMF reference spectrum. This interference pattern, also visible in the raw data (not co-added spectra of the SMF outputs), is suspected to result from internal reflections on the PL and VG end faces (due to the lack of an AR coating) having a Fabry-Perot effect, but requires further investigation.

Finally, the spectra reconstructed from the PL data show a decrease in intensity for wavelengths larger than about 760~nm, plus a hump at about 775~nm. This is due to the 800~nm short pass filter used at an angle for the pyramid WFS pickoff.  The angle of the filter relative to the incident beam influences the transmission of both parallel and perpendicular polarizations of light in varying ways. The result is a bump located here around 770~nm in the spectral data, which is also visible in the SMF spectrum.

\subsubsection{Injection efficiency versus focal ratio}
\label{sec:injeff}

Our procedure was similar for each tested focal ratio: we optimized the injection by acquiring a coupling map with the SCExAO \rev{SuperK} and positioning the PL at the estimated center of the coupling map. \rev{Because our SuperK is not very bright in the visible wavelength, we needed to use a more powerful visible laser in order to make flux measurements using a power meter placed in the beam before and after the PL. After optimization on the SuperK, we swapped the source} for a 642~nm laser, and recorded the flux intensity, with the power meter, before ($I_b$) and after ($I_a$) the PL. We normalized out the Fresnel reflections ($f_r$) at the input and output of the PL ($96\%$ per surface), plus the average throughput ($t_p$) of the PL ($88\%$) in order to compute the coupling injection efficiency ($c_\text{eff}$), following $c_\text{eff} = I_a / (I_b \times f_{r}^2 \times t_p)$. Owing to the granulation in the coupling maps, the coupling efficiency might not be the same over the whole FoV. Using the on-axis measurement, we converted the coupling map computed at $642$~nm into a coupling efficiency map by extrapolating the on-axis $c_\text{eff}$ value to the rest of the map. From this coupling efficiency map, we computed two more quantities: the maximum coupling efficiency and the averaged injection efficiency over the FoV. We chose to define the FoV as the angular separation for which the injected flux in the lantern is half of the normalized flux in Fig.~\ref{fig:PL_coupling_wl}.

Figure~\ref{fig:PL_coupling} summarizes our tests, highlighting the on-axis, maximum, and averaged injection efficiency for various focal ratios. The error bars of the averaged injection efficiency represent the standard deviation of the coupling efficiency over the FoV. The injection efficiency shows a relatively fast increase from focal ratios 4 to 8. The highest injection efficiency is obtained for the focal ratio of 8, with a\rev{n average} value of $51\% \pm 10\%$ over an FoV of about \rev{80}~mas\rev{, an on-axis injection efficiency of $61.8\%$,} and a maximum injection efficiency of $80\%$. The injection efficiency then decreases for focal ratios ranging from 8 to 28. This trend was observed in \cite{lin2022experimental} when studying the coupling efficiency into a 19-port PL optimized for near-infrared wavelengths.
The graph also shows the FoV evolution as a function of the focal ratio. The curve follows a hyperbola. This is expected since the acceptance angle of a fiber is proportional to the inverse of the focal ratio. 

This graph can be used to assess which PL injection focal ratio to use depending on the scientific application. Lower focal ratios provide a larger FoV but with limited injection efficiency. Additionally, as we have seen previously, there are spectral information losses that would prevent optimal continuum reconstruction due to modal overfilling of the PL. On the other hand, higher focal ratios offer a smaller FoV but provide better spectral reconstruction.

\subsubsection{Injection efficiency versus Strehl ratio}
\label{sec:injeffstehl}

To assess the injection efficiency into the PL in the presence of uncorrected atmospheric turbulence, turbulence is injected onto the SCExAO DM with various levels of upstream atmospheric turbulence correction. Turbulence screens are based on Kolmogorov spectrum with inner and outer scales, adopting the frozen flow approximation for temporal evolution\rev{, and are the closest to what we expect during on-sky observations}. We fixed the wind speed at 10 m/s and modified the turbulence amplitude to vary the Strehl ratio. In order to study the behavior of the PL in various conditions, we varied the spatial frequency content of the simulated turbulence. We identified three distinct cases by adjusting the inner and outer scales of the turbulence. \rev{First,} turbulence following the Kolmogorov power spectrum (inner scale = 0.01 meter, outer scale = 20 meters)\rev{, which is the closest to what we would commonly experience on SCExAO}. \rev{Second,} turbulence dominated by high spatial frequencies (inner scale = 0.01 meter, outer scale = 1 meter)\rev{, which would correspond to an ExAO case where low-order aberrations are well corrected but there are still remaining high-order aberrations (for example, if the DM lacks actuators)}. \rev{Third,} turbulence dominated by the low spatial frequencies (inner scale = 10 meters, outer scale = 100 meters)\rev{, which would correspond to a more classic AO loop performing poorly on low-order aberrations (for example, during the observation of a faint star).}

We chose to conduct this study with the injection set at a focal ratio of about 8, where our on-axis injection efficiency was \rev{$61.8\%$ at 642~nm}. The result is presented in Fig.~\ref{fig:PL_injeffstrehl}. The Strehl ratio was estimated from focal plane images simultaneously recorded on the VAMPIRES instrument at 750~nm. Our reference point was defined as the condition where there is no simulated turbulence on the bench. 

The estimated Strehl ratio without simulated turbulence was \rev{estimated at} $90\%$ \rev{from the VAMPIRES images}, which we associated with the $61.8\%$ injection efficiency measured on the bench. For each tested Strehl ratio, a total of 1000 \rev{PL} frames were acquired. We performed \rev{the} spectrum reconstruction from each frame \rev{following the same process than described in Sect.~\ref{sec-injvswl}. We computed the average and standard deviation of the flux over the 1000 frames for each spectral channel.} We then compared the averaged values \rev{for the 642~nm spectral channel} to our reference to estimate the injection efficiency. Additionally, we mapped the evolution of the injection efficiency as a function of Strehl ratio for a simulated SMF. This simulation was conducted by \cite{lin2021design} in the context of designing PLs for ground-based telescopes and, among other aspects, involved comparing their theoretical injection efficiency with that of an SMF. If the simulations were run based on synthetic images of the Keck telescope, the results could easily be scaled to the Subaru Telescope. A study by \cite{jovanovic2017efficient} demonstrated that the injection efficiency into an SMF on the Subaru Telescope should be $92.3\%$ of that on the Keck telescope, due to the change in the pupil shape. The graph also displays the various turbulence regimes (seeing, AO, and ExAO).

The graph indicates that the injection efficiency into the PL exhibits a linear relationship with the Strehl ratio across all turbulence regimes in the case in which high-frequency contents are favored in the turbulence simulation. This behavior is similar to the one of the SMF\rev{, and can be explained by the non-sensitivity of the PL (or the SMF) to features at large separations}. When the turbulence is dominated by low-frequency aberrations, the injection to Strehl relationship remains linear but exhibits a slower decrease in efficiency with the Strehl degradation compared to the high-frequency dominated case. In the regime in which the turbulence follows a classic Kolmogorov spectrum, the linearity is similar to that observed with low-frequency trends for both AO and ExAO turbulence regimes, until there is a drop in the seeing regime. This latter decline is attributed to tip-tilt dominating aberrations, as is explained and simulated in \cite{lin2021design}.

This study reveals that the PL behaves similarly to a SMF, in terms of injection efficiency, when high-order aberrations dominate. In a turbulence regime dominated by lower-spatial-frequency aberrations, the injection efficiency decreases slower with Strehl ratio degradation. In this latter regime, a PL provides higher coupling efficiency than an SMF, especially for Strehl ratios below 0.8. In practice, a PL and an SMF will have similar behaviors in the ExAO regime - with an advantage given to the PL \rev{especially} in the presence of small residual tip or tilt. In the AO regime, a PL will have a better injection efficiency compared to an SMF.

\section{On-sky demonstration }
\label{PL-onsky}

\subsection{Observation log}

The focal ratio selected for this observation was~8, providing the best injection efficiency. We observed Ikiiki ($\alpha$ Leo, $m_R$ = 1.37) and `Aua ($\alpha$ Ori, $m_R$ = -1.17) on February 18, 2024 UTC during an engineering night (Proposal ID: S24A-EN01, PI : Julien Lozi, Support Astronomers: Julien Lozi and  Sébastien Vievard, Telescope Operator : Erin Dailey). Data was acquired at a 200~Hz framerate (see Table~\ref{table:Log_obs}). The seeing in the H band was estimated to be about $1$\arcsec{} by the SCExAO NIR internal fast frame camera. \rev{The VAMPIRES focal plane camera was used to estimate the residual tip or tilt and the Strehl in the visible, during the observation of Ikiiki, at 670~nm, 720~nm and 760~nm. The tip-tilt residual was about 12~mas RMS across all wavelengths. The estimated Strehl ratio was on average $26\%$ at 670~nm, $28\%$ at 720~nm, and $30\%$ at 760~nm.}

\begin{table}[h!]
	\centering
	\caption{Observation log.}
	\label{table:Log_obs}
	\begin{tabular}{ccccc}
		\hline 
		Target                      &  $N_{img}$     & $t_{int} (ms)$    &   Seeing  & \rev{Strehl estimate}\\
		\hline 
		
		Ikiiki						& 	240,000 	& 	5 			&  1\arcsec{} & 
            { \begin{tabular}{c}
                            \rev{$25.9\%$ (at 670~nm)} \\ 
                            \rev{$28.1\%$ (at 720~nm)} \\ 
                            \rev{$30.4\%$ (at 760~nm)}
                            \end{tabular}} \\
            \hline
		`Aua 					&   636,000 	&   5 			&  1\arcsec{} & \rev{N/A} \\
		
		\hline 
		\end{tabular}
\end{table}

\subsection{Observation of Ikiiki}

We observed Ikiiki for 20~minutes at a 200~Hz frame rate, providing a total of 240,000 frames. Because the observing conditions were not particularly good, not all of the acquired frames could be exploited. We selected the best $90~\%$ of the frames based on the flux received by the PL science camera. Figure~\ref{fig:Reg-img} shows a dark-subtracted averaged image after frame selection. We can already identify, in this data, three prominent absorption lines. We also observe that the flux distribution varies across traces, indicating the sensitivity of PL to the input scene or wavefront. We calculated the S/N for each minute of observation by dividing the average flux by the standard deviation of the flux, adjusted for the number of frames (i.e., the error on the flux measurement), per spectral channel over time. This approach yielded S/N values that varied around 200, primarily due to fluctuations in the injection efficiency.

\setcounter{figure}{15-1}
\begin{figure}[h]
        \centering
        \includegraphics[width=\linewidth]{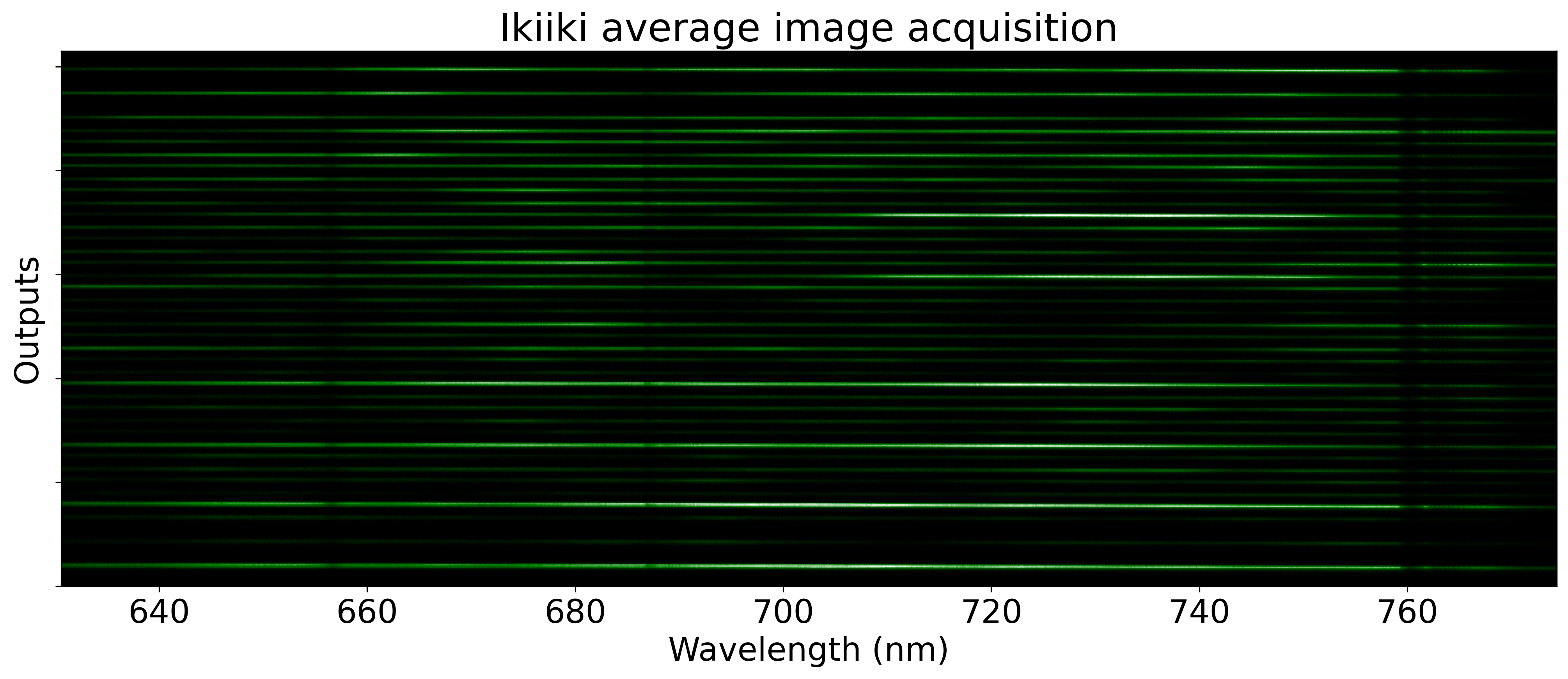}
        \caption{Spectra of Ikiiki acquired with the PL. The horizontal axis is the wavelength, and the 38 output traces corresponding to the two polarizations from each of the PL 19 SMF outputs are stacked along the vertical axis. This image was computed from 20 minutes of observation at a 200~Hz frame rate, with the best $90~\%$ frames selected.}
        \label{fig:Reg-img}
\end{figure}{}

\setcounter{figure}{16-1}
\begin{figure*}[t]
\sidecaption
        \centering
        \includegraphics[width=12cm]{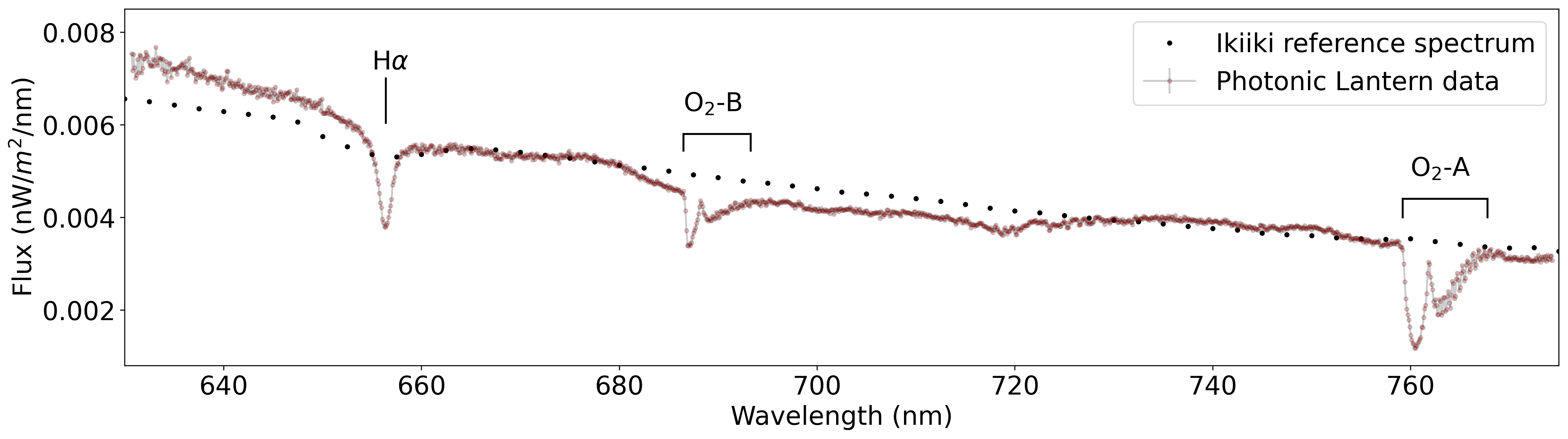}
        \caption{Ikiiki calibrated spectrum compared to its reference. The spectral features detected are oxygen bands A and B from the Earth's atmosphere (around 761~nm and 687~nm respectively), and H$\alpha$ (at about 656~nm).}
        \label{fig:Reg-spectrum}
\end{figure*}{}

\subsubsection{Spectrum reconstruction}
\label{reg_spectr_S/N}

For each selected frame, we subtracted the dark+bias and extracted each PL output trace. The traces were then co-added for each frame, and then averaged over the entire observation sequence. 

To calibrate the response of our instrument, we used a flat-field halogen lamp (approximately black body spectrum) positioned at the entrance of AO188 to uniformly illuminate the PL input FoV. By performing the same computation as was previously described, we derived a flat field spectrum, which we then calibrated using the theoretical black body slope corresponding to the Halogen lamp temperature (3020.3K). Subsequently, we utilized the resulting calibrated flat data to calibrate the Ikiiki reconstructed spectrum obtained with the PL. We further refined the computed spectrum by normalizing it to match a reference spectrum obtained from~\cite{1996BaltA...5..603A}, represented by black dots in Fig.~\ref{fig:Reg-spectrum}. 

\setcounter{figure}{17-1}
\begin{figure*}[h]
        \centering
 \sidecaption
        \begin{tabular}{c}
            \includegraphics[width=12cm]{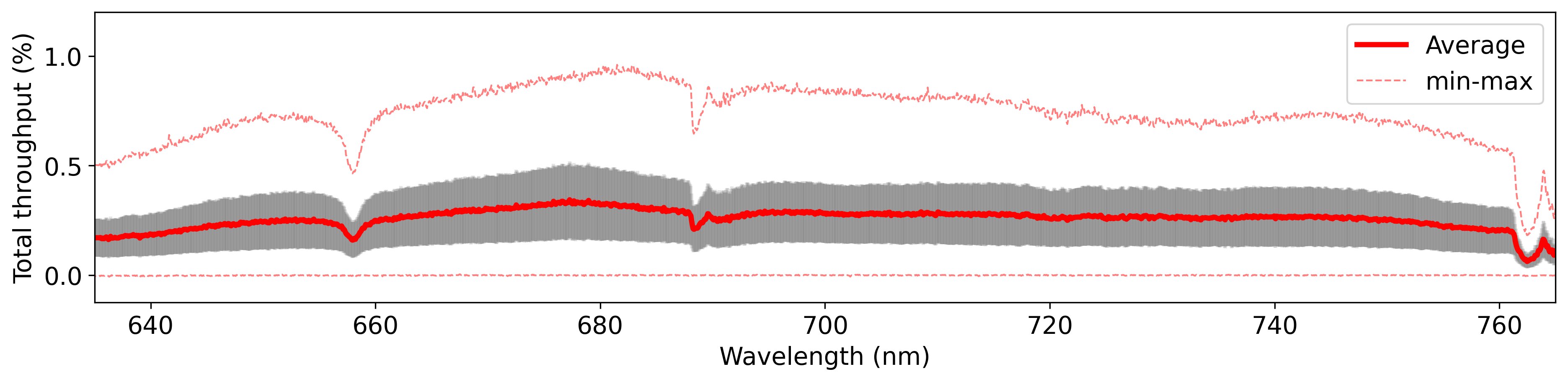} \\
            \includegraphics[width=12cm]{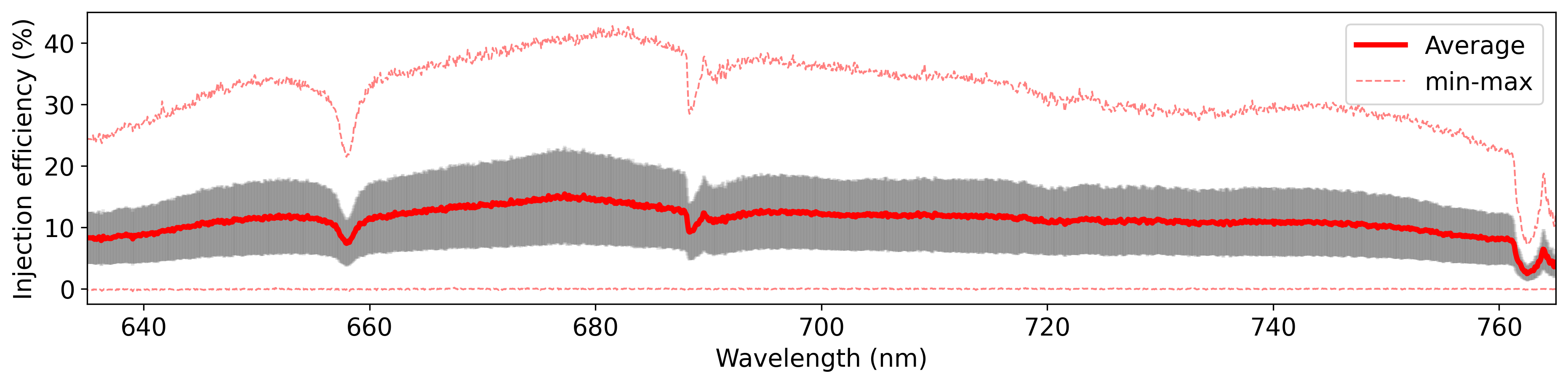}\\ 
        \end{tabular}
        \caption{\rev{Total throughput of the instrument (top panel) and injection efficiency into the PL (bottom panel) as a function of the wavelength, during the observation of Ikiiki. The average values are plotted in thick red, along with the standard deviation. The minimum and maximum throughput is also plotted in dashed red lines.}}
        \label{fig:onky_effi}
\end{figure*}{}

On the same graph, we show the resulting calibrated and adjusted spectrum of Ikiiki obtained with the PL. The latter closely matches the reference spectrum. Some differences can be noted, which may be attributed to the imperfect calibration of our data. Indeed, calibrating our reconstructed spectrum with the halogen lamp may potentially overlook chromatic perturbations caused by the atmosphere and/or various optical surfaces prior to AO188. Ideally, calibration should be conducted using a well-calibrated point source in the sky. The spectrum of Ikiiki exhibits three distinct absorption lines: the atmospheric oxygen A and B bands (around 761~nm and 687~nm, respectively), and H$\alpha$ (at approximately 656~nm). The absorption in the two oxygen bands is due to oxygen molecules in Earth's atmosphere and does not appear in the reference spectrum due to its terrestrial origin. At a spectral resolution of R$\approx$3000, the $O_2$ absorption bands appear as a deep broad absorption line with a forest of narrower lines at longer wavelengths. The H$\alpha$ absorption line originates from the cooler outer layers of the Ikiiki atmosphere, where neutral hydrogen atoms absorb photons emitted by the deeper, hotter layers at the energy level corresponding to the H$\alpha$ transition.

\subsubsection{Injection efficiency}
In order to estimate the on-sky injection efficiency of the PL, we compared the acquired spectra (prior to flat-field calibration) to the reference spectro-photometric data of Ikiiki~\citep{1996BaltA...5..603A}. We summarize in Table~\ref{tab:tel_throughput} the throughput of  elements from the entrance of the telescope down to the detector. We considered the detector quantum efficiency provided by the manufacturer (from $78\%$ to $54\%$ from 600~nm to 785~nm). We also modeled the throughput of the Earth's atmosphere using the synthetic photometry Python package (synphot) maintained by the Space Telescope Science Institute (STScI). The airmass during the observation was 1.01, and induced an extinction ranging from 0.1 to 0.03 magnitude between 600~nm and 770~nm, leading to a throughput of on average about $95\%$.

\begin{table}[!h]
    \centering
    \caption{Throughput of the various optical elements from the entrance of the telescope to the detector.}
    \begin{tabular}{cc}
    \hline
         Optical element & Throughput \\
    \hline
         Subaru Telescope (M1-M2-M3) & \rev{$70\%$}   \\
         \rev{AO 188} & \rev{$65\%$} \\
         \rev{SCExAO PTTB} & \rev{$58\%$} \\
         PyWFS filter & $86\%$ \\
         Beamsplitter Cube & $40\%$ \\
         Injection lenses (x3) & $99\%$ \\
         Photonic Lantern & $88\%$ \\
         Objective & $96\%$ \\
         Wollaston & $80\%$ \\
         VPH & $70\%$ \\
         Achromatic Doublet (x2) & $99\%$ \\
         \hline
    \end{tabular}
     \tablefoot{\rev{Subaru Telescope M1-M2-M3 stands for the three mirrors before the light enters AO188. SCExAO PTTB (Path to Top Bench) is from the SCExAO entrance to the output of the periscope, from ~\cite{2015PASP..127..890J}.} \rev{Values are provided for R-band.}}
    \label{tab:tel_throughput}
\end{table}

\setcounter{figure}{20-1}
\begin{figure*}[!b]
        \sidecaption
	\centering
	\includegraphics[width=12cm]{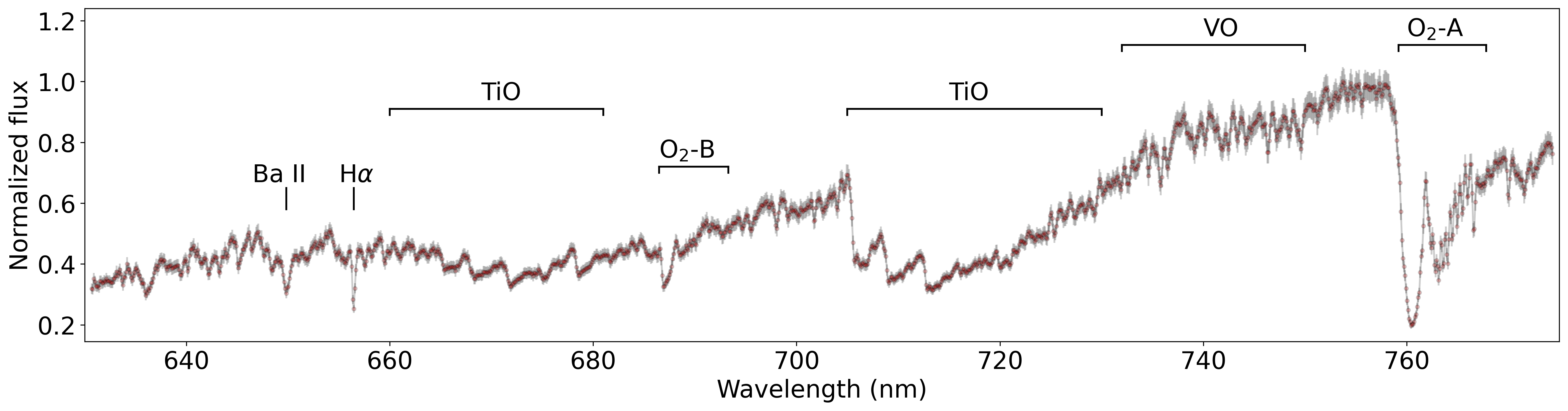}
	\caption{'Aua calibrated and normalized spectrum. Multiple molecular aborption bands are observed in the spectrum of this red supergiant star. Terrestrial atmosphere $O_2$ bands are also retrieved.}
	\label{fig:Bet-spectrum}
\end{figure*}{}

\rev{We calculated the total instrument throughput and the injection efficiency for each frame and each spectral channel. Figure~\ref{fig:onky_effi} shows the resulting average values, standard deviation and min-max values for each spectral channel. At 642~nm (see top panel of Figure~\ref{fig:throughp_combined}), we report a total average throughput of the instrument of $0.2\pm0.1\%$ with a maximum of $0.6\%$. We report an average injection efficiency of $9.4\pm 4.9\%$ with a maximum of $28.0\%$. The best injection efficiency performance at 642~nm is about half of that obtained on the bench under optimal conditions. This is in accordance with  Figure~\ref{fig:PL_injeffstrehl} which predicted an injection efficiency no greater than $30\%$ for a Strehl below $25\%$. We notice that the injection efficiency drops to zero. This can be explained by poor performance of the AO loops under the 1" seeing estimated in H band. Thanks to simultaneous imaging using the VAMPIRES instrument, we were able to compute a residual tip/tilt larger than 10~mas RMS in average during the observation of Ikiiki, contributing to degrading the injection efficiency into the PL. The best performance obtained was around 680~nm (see bottom panel of Figure~\ref{fig:throughp_combined}), where we report an average total throughput of $0.3\pm0.2\%$ with a maximum of $0.9\%$, and an average injection efficiency of $14.5\pm 7.4\%$ with a maximum of $42.8\%$.} 

\setcounter{figure}{18-1}
\begin{figure}[!h]
	\centering
	\includegraphics[width=\linewidth]{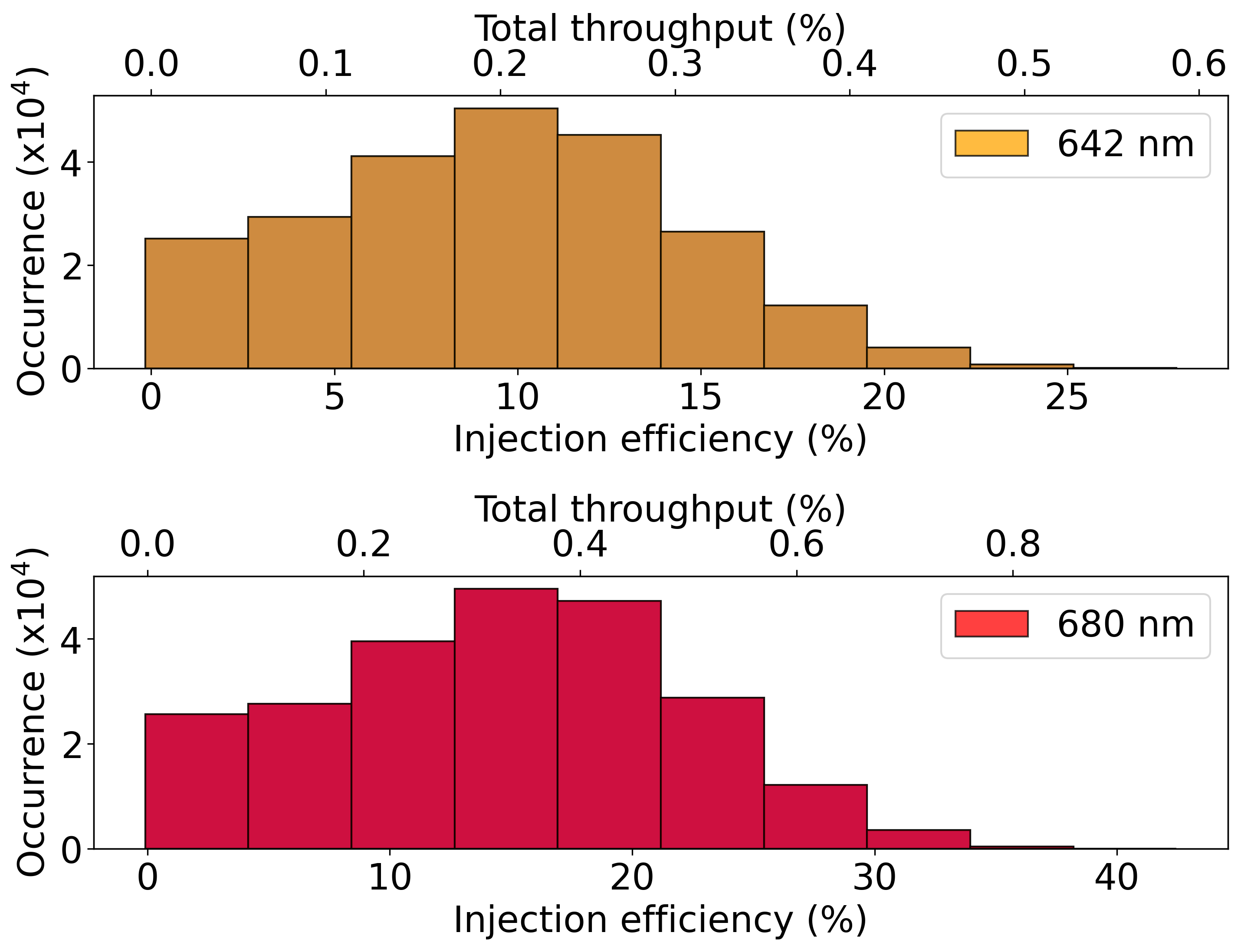}\\
 \caption{\rev{Distributions of total throughput and injection efficiency at 642~nm (top) and 680~nm (bottom) during the observation of Ikiiki.}}
	\label{fig:throughp_combined}
\end{figure}{}

Even under these difficult conditions, the collected data had a good S/N and enabled the faithful reconstruction of the Ikiiki spectrum.



\subsection{Observation of `Aua}

We observed `Aua for 53~minutes at a 200~Hz frame rate, providing a total of 636,000 frames. Similarly to the previous target, we performed a frame selection, since the conditions were not good. The best $90\%$ of frames were selected, dark+bias subtracted, and averaged to compute the image displayed in Fig.~\ref{fig:Bet-img}. Many spectral features can be identified on the averaged image. The computed S/N per minute was around 130, a sign that the observation conditions worsened compared to the previous target (where the S/N per minute was 200 in average).

\setcounter{figure}{19-1}
\begin{figure}[!h]
        \centering
        \includegraphics[width=\linewidth]{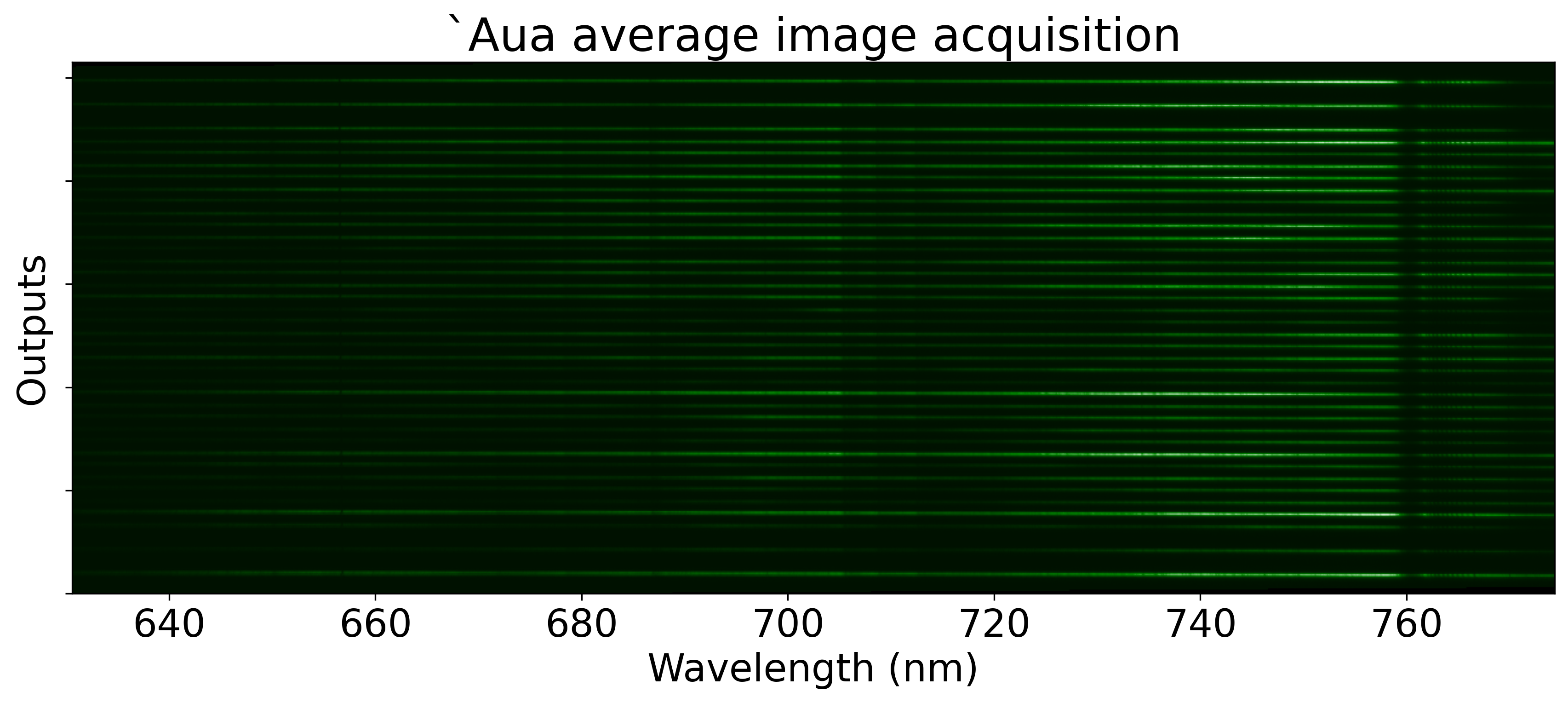}
        \caption{Spectra of `Aua acquired with the PL. The horizontal axis is the wavelength dispersion, and vertically we have the 38 output traces corresponding to the two polarizations from each of the PL 19 SMF outputs. This image was computed from 53~minutes of observation at a framerate of 200 Hz, after selection of the best $90\%$ of frames.}
        \label{fig:Bet-img}
\end{figure}{}

After trace extraction and calibration with flat-field data, we co-added the resulting spectra to obtain the spectrum shown in Fig.~\ref{fig:Bet-spectrum}. The spectrum shows multiple absorption spectral lines. The absorption lines detected on Ikiiki are present: oxygen bands A and B from the Earth's atmosphere around 761~nm and 687~nm, respectively, and H$\alpha$ absorption by the stellar atmosphere at about 656~nm. Many titanium oxide (TiO) lines are also present in the spectrum, with the strongest at about 705~nm. TiO lines in red supergiants are characteristic of convection cells within the star's atmosphere~\citep{10.1117/12.459135,refId0}, and are well known in the case of 'Aua as they are extensively exploited for image reconstruction using interferometry~\citep{2000MNRAS.315..635Y,levesque2005effective,chiavassa2010radiative}. The ability to detect these (and the following) spectral lines, while conserving spatial information via the use of a PL opens the door to future use of differential phase interferometry~\citep{Akeson2000} to recover spatial features (such as stellar surface details) from PL observations.
We can also spot the presence of vanadium oxide (VO) lines from about 730~nm to 750~nm. Similarly to TiO lines, VO lines are generated by convection cells in red supergiant stars, although they are less dominant than TiO. Finally, we identified the barium II (Ba II) line at about 649~nm~\citep{koza2010sensitivity}. These lines can also be seen in data from the X-shooter Spectral Library (XSL)~\citep{gonneau2020x} taken at the VLT-UT2 using a medium-resolution spectrograph. This demonstration highlights the capabilities of the PL signal reconstruction to retrieve a target's spectral features. The next step, not addressed in this demonstration paper, will be to exploit the spectral reconstruction to extract astronomical information about the observed target.

\section{Conclusion}

We have presented the integration, characterization, and on-sky demonstration of a PL at the 8.2~m Subaru Telescope. Photonic lanterns allow for high-efficiency SMF-based spectroscopy thanks to their unique design. Their MM input insures high coupling efficiency, compared to a single SMF, especially at visible wavelengths at which even ExAO systems struggle to reach high Strehl ratios. The SM outputs of PLs are perfectly suited for spectroscopy using compact and stable instruments.

Depending on the optical injection setup focal ratio, the PL can provide various FoVs and injection efficiency performance depending on the desired science output. Our setup shows an optimal injection efficiency for a focal ratio of about 8, providing an injection efficiency of about $51\% \pm 10\%$ over an FoV of \rev{80}~mas, with an efficiency peak of $80\%$ \rev{at 642~nm}. 

We studied the relationship between the PL injection efficiency and Strehl ratio. Various turbulence profiles were simulated on the SCExAO DM, and the results were confronted to simulations in the case of an SMF. We have shown that PLs have a similar linear relationship between the injection efficiency and the Strehl to that of an SMF. However, PLs perform better in the presence of turbulence dominated by low-frequency aberrations.

Finally, we tested the PL on-sky. Two stars were observed: a point source, Ikiiki ($\alpha$ Leo), and a red supergiant star, `Aua ($\alpha$ Ori). We were first able to reconstruct Ikiiki's spectrum with high fidelity compared to its reference. Under median seeing conditions, the injection efficiency into the PL averaged to \rev{$9.4\% \pm 4.9\%$ and reached up to $28\%$ at 642~nm, leading to an instrument throughput of up to $0.6\%$. The best performance was obtained at 680~nm, where the injection efficiency into the PL averaged $14.5\% \pm 7.4\%$ and reached up to $42.8\%$, leading to an instrument throughput of up to $0.9\%$. It is important to note that these figures likely fall short of demonstrating the full capabilities of our visible PL under more favorable conditions. For context, attempting a similar light injection directly into a single SMF under these conditions would likely yield inferior results.} The observation of `Aua confirms our ability to reconstruct a complex spectrum from the PL data. We were able to retrieve multiple spectral features characteristic of this target.

\rev{The on-sky} performance should be increased under better atmospheric conditions, and even further with the soon-to-arrive upgrades of the Subaru AO facility DM (from 188~elements to 3000~elements) and visible wavefront sensor~\citep{ahn2023non}. \rev{We also plan on upgrading the NPBS, picking up the light from SCExaO, and routing it to the PL with a R90/T10 BS. This should significantly increase the overall throughput of the instrument.}


This result shows that our setup can be used for high-throughput SMF-fed spectroscopy in the visible, especially for H$\alpha$ spectro-imaging. The ongoing and near-future work aims to provide this capability to the community as a feature of the FIRST instrument. This demonstration constitutes the first on-sky acquisition of a spectrum using a PL fed by an ExAO system in the visible.

\begin{acknowledgements}
The development of SCExAO was supported by the Japan Society for the Promotion of Science (Grant-in-Aid for Research \#23340051, \#26220704, \#23103002, \#19H00703 \& \#19H00695), the Astrobiology Center of the National Institutes of Natural Sciences, Japan, the Mt Cuba Foundation and the director’s contingency fund at Subaru Telescope. This work was supported by the National Science Foundation under Grant No. 2109231.
K.A. acknowledges funding from the Heising-Simons foundation.
V.D. acknowledges support from NASA funding (Grant \#80NSSC19K0336).
M.T. is supported by JSPS KAKENHI grant Nos.18H05442, 15H02063, and 22000005.
\manon{M.L acknowledges support from the doctoral school Astronomy and Astrophysics of Ile de France (ED 127) and \manon{from the French National Research Agency (ANR-21-CE31-0005).}}
S.L. also acknowledges the support of the ANR, under grants ANR-21-CE31-0017 (project ExoVLTI) and ANR-22-EXOR-0005 (PEPR Origins).
The authors wish to recognize and acknowledge the very significant cultural role and reverence that the summit of Mauna Kea has always had within indigenous Hawaiian communities, and are most fortunate to have the opportunity to conduct observations from this mountain.
\end{acknowledgements}

\bibliographystyle{bibtex/aa}
\bibliography{main}

\begin{thebibliography}{54}
\expandafter\ifx\csname natexlab\endcsname\relax\def\natexlab#1{#1}\fi

\bibitem[{Afram \& Berdyugina(2019)}]{refId0}
Afram, N. \& Berdyugina, S.~V. 2019, Astronomy \& Astrophysics, 629, A83

\bibitem[{Ahn {et~al.}(2023)Ahn, Guyon, Lozi, Vievard, Deo, Lallement, \&
  Bragg}]{ahn2023non}
Ahn, K., Guyon, O., Lozi, J., {et~al.} 2023, in Techniques and Instrumentation
  for Detection of Exoplanets XI, Vol. 12680, SPIE, 75--83

\bibitem[{{Akeson} {et~al.}(2000){Akeson}, {Swain}, \& {Colavita}}]{Akeson2000}
{Akeson}, R.~L., {Swain}, M.~R., \& {Colavita}, M.~M. 2000, in Society of
  Photo-Optical Instrumentation Engineers (SPIE) Conference Series, Vol. 4006,
  Interferometry in Optical Astronomy, ed. P.~{L{\'e}na} \& A.~{Quirrenbach},
  321--327

\bibitem[{{Alekseeva} {et~al.}(1996){Alekseeva}, {Arkharov}, {Galkin},
  {Hagen-Thorn}, {Nikanorova}, {Novikov}, {Novopashenny}, {Pakhomov}, {Ruban},
  \& {Shchegolev}}]{1996BaltA...5..603A}
{Alekseeva}, G.~A., {Arkharov}, A.~A., {Galkin}, V.~D., {et~al.} 1996, Baltic
  Astronomy, 5, 603

\bibitem[{Birks {et~al.}(2015)Birks, Gris-S{\'a}nchez, Yerolatsitis,
  Leon-Saval, \& Thomson}]{birks2015photonic}
Birks, T.~A., Gris-S{\'a}nchez, I., Yerolatsitis, S., Leon-Saval, S., \&
  Thomson, R.~R. 2015, Advances in Optics and Photonics, 7, 107

\bibitem[{{Boogert} {et~al.}(2002){Boogert}, {Hogerheijde}, \&
  {Blake}}]{boogert02}
{Boogert}, A.~C.~A., {Hogerheijde}, M.~R., \& {Blake}, G.~A. 2002, \apj, 568,
  761

\bibitem[{{Brogi} {et~al.}(2012){Brogi}, {Snellen}, {de Kok}, {Albrecht},
  {Birkby}, \& {de Mooij}}]{brogi12}
{Brogi}, M., {Snellen}, I. A.~G., {de Kok}, R.~J., {et~al.} 2012, \nat, 486,
  502

\bibitem[{{Casares} \& {Jonker}(2014)}]{casares14}
{Casares}, J. \& {Jonker}, P.~G. 2014, \ssr, 183, 223

\bibitem[{Chiavassa {et~al.}(2010)Chiavassa, Haubois, Young, Plez, Josselin,
  Perrin, \& Freytag}]{chiavassa2010radiative}
Chiavassa, A., Haubois, X., Young, J., {et~al.} 2010, Astronomy \&
  Astrophysics, 515, A12

\bibitem[{{Conrad} {et~al.}(2015){Conrad}, {de Kleer}, {Leisenring}, {La
  Camera}, {Arcidiacono}, {Bertero}, {Boccacci}, {Defr{\`e}re}, {de Pater},
  {Hinz}, {Hofmann}, {K{\"u}rster}, {Rathbun}, {Schertl}, {Skemer},
  {Skrutskie}, {Spencer}, {Veillet}, {Weigelt}, \& {Woodward}}]{conrad15}
{Conrad}, A., {de Kleer}, K., {Leisenring}, J., {et~al.} 2015, \aj, 149, 175

\bibitem[{{Currie} {et~al.}(2023{\natexlab{a}}){Currie}, {Biller}, {Lagrange},
  {Marois}, {Guyon}, {Nielsen}, {Bonnefoy}, \& {De Rosa}}]{currie2023ppvii}
{Currie}, T., {Biller}, B., {Lagrange}, A., {et~al.} 2023{\natexlab{a}}, in
  Astronomical Society of the Pacific Conference Series, Vol. 534, Protostars
  and Planets VII, ed. S.~{Inutsuka}, Y.~{Aikawa}, T.~{Muto}, K.~{Tomida}, \&
  M.~{Tamura}, 799

\bibitem[{{Currie} {et~al.}(2023{\natexlab{b}}){Currie}, {Brandt}, {Brandt},
  {Lacy}, {Burrows}, {Guyon}, {Tamura}, {Liu}, {Sagynbayeva}, {Tobin},
  {Chilcote}, {Groff}, {Marois}, {Thompson}, {Murphy}, {Kuzuhara}, {Lawson},
  {Lozi}, {Deo}, {Vievard}, {Skaf}, {Uyama}, {Jovanovic}, {Martinache},
  {Kasdin}, {Kudo}, {McElwain}, {Janson}, {Wisniewski}, {Hodapp}, {Nishikawa},
  {He{\l}miniak}, {Kwon}, \& {Hayashi}}]{currie2023}
{Currie}, T., {Brandt}, G.~M., {Brandt}, T.~D., {et~al.} 2023{\natexlab{b}},
  Science, 380, 198

\bibitem[{{Currie} {et~al.}(2022){Currie}, {Lawson}, {Schneider}, {Lyra},
  {Wisniewski}, {Grady}, {Guyon}, {Tamura}, {Kotani}, {Kawahara}, {Brandt},
  {Uyama}, {Muto}, {Dong}, {Kudo}, {Hashimoto}, {Fukagawa}, {Wagner}, {Lozi},
  {Chilcote}, {Tobin}, {Groff}, {Ward-Duong}, {Januszewski}, {Norris},
  {Tuthill}, {van der Marel}, {Sitko}, {Deo}, {Vievard}, {Jovanovic},
  {Martinache}, \& {Skaf}}]{currie2022}
{Currie}, T., {Lawson}, K., {Schneider}, G., {et~al.} 2022, Nature Astronomy,
  6, 751

\bibitem[{Cvetojevic {et~al.}(2012)Cvetojevic, Jovanovic, Betters, Lawrence,
  Ellis, Robertson, \& Bland-Hawthorn}]{cvetojevic2012first}
Cvetojevic, N., Jovanovic, N., Betters, C., {et~al.} 2012, Astronomy \&
  Astrophysics, 544, L1

\bibitem[{Delorme {et~al.}(2021)Delorme, Jovanovic, Echeverri, Mawet,
  Kent~Wallace, Bartos, Cetre, Wizinowich, Ragland, Lilley,
  {et~al.}}]{delorme2021keck}
Delorme, J.-R., Jovanovic, N., Echeverri, D., {et~al.} 2021, Journal of
  Astronomical Telescopes, Instruments, and Systems, 7, 035006

\bibitem[{{Do} {et~al.}(2019){Do}, {Hees}, {Ghez}, {Martinez}, {Chu}, {Jia},
  {Sakai}, {Lu}, {Gautam}, {O'Neil}, {Becklin}, {Morris}, {Matthews},
  {Nishiyama}, {Campbell}, {Chappell}, {Chen}, {Ciurlo}, {Dehghanfar},
  {Gallego-Cano}, {Kerzendorf}, {Lyke}, {Naoz}, {Saida}, {Sch{\"o}del},
  {Takahashi}, {Takamori}, {Witzel}, \& {Wizinowich}}]{do19}
{Do}, T., {Hees}, A., {Ghez}, A., {et~al.} 2019, Science, 365, 664

\bibitem[{{Fernandes} {et~al.}(2019){Fernandes}, {Mulders}, {Pascucci},
  {Mordasini}, \& {Emsenhuber}}]{fernandes19}
{Fernandes}, R.~B., {Mulders}, G.~D., {Pascucci}, I., {Mordasini}, C., \&
  {Emsenhuber}, A. 2019, \apj, 874, 81

\bibitem[{{Ford} {et~al.}(2014){Ford}, {McKernan}, {Sivaramakrishnan},
  {Martel}, {Koekemoer}, {Lafreni{\`e}re}, \& {Parmentier}}]{ford14}
{Ford}, K.~E.~S., {McKernan}, B., {Sivaramakrishnan}, A., {et~al.} 2014, \apj,
  783, 73

\bibitem[{Gonneau {et~al.}(2020)Gonneau, Lyubenova, Lan{\c{c}}on, Trager,
  Peletier, Arentsen, Chen, Coelho, Dries, Falc{\'o}n-Barroso,
  {et~al.}}]{gonneau2020x}
Gonneau, A., Lyubenova, M., Lan{\c{c}}on, A., {et~al.} 2020, Astronomy \&
  Astrophysics, 634, A133

\bibitem[{{Haffert} {et~al.}(2019){Haffert}, {Bohn}, {de Boer}, {Snellen},
  {Brinchmann}, {Girard}, {Keller}, \& {Bacon}}]{haffert19}
{Haffert}, S.~Y., {Bohn}, A.~J., {de Boer}, J., {et~al.} 2019, Nature
  Astronomy, 3, 749

\bibitem[{{Jovanovic} {et~al.}(2015){Jovanovic}, {Martinache}, {Guyon},
  {Clergeon}, {Singh}, {Kudo}, {Garrel}, {Newman}, {Doughty}, {Lozi}, {Males},
  {Minowa}, {Hayano}, {Takato}, {Morino}, {Kuhn}, {Serabyn}, {Norris},
  {Tuthill}, {Schworer}, {Stewart}, {Close}, {Huby}, {Perrin}, {Lacour},
  {Gauchet}, {Vievard}, {Murakami}, {Oshiyama}, {Baba}, {Matsuo}, {Nishikawa},
  {Tamura}, {Lai}, {Marchis}, {Duchene}, {Kotani}, \&
  {Woillez}}]{2015PASP..127..890J}
{Jovanovic}, N., {Martinache}, F., {Guyon}, O., {et~al.} 2015, Publications of
  the Astronomical Society of the Pacific, 127, 890

\bibitem[{Jovanovic {et~al.}(2017)Jovanovic, Schwab, Guyon, Lozi, Cvetojevic,
  Martinache, Leon-Saval, Norris, Gross, Doughty,
  {et~al.}}]{jovanovic2017efficient}
Jovanovic, N., Schwab, C., Guyon, O., {et~al.} 2017, Astronomy \& Astrophysics,
  604, A122

\bibitem[{{Keppler} {et~al.}(2018){Keppler}, {Benisty}, {M{\"u}ller},
  {Henning}, {van Boekel}, {Cantalloube}, {Ginski}, {van Holstein}, {Maire},
  {Pohl}, {Samland}, {Avenhaus}, {Baudino}, {Boccaletti}, {de Boer},
  {Bonnefoy}, {Chauvin}, {Desidera}, {Langlois}, {Lazzoni}, {Marleau},
  {Mordasini}, {Pawellek}, {Stolker}, {Vigan}, {Zurlo}, {Birnstiel},
  {Brandner}, {Feldt}, {Flock}, {Girard}, {Gratton}, {Hagelberg}, {Isella},
  {Janson}, {Juhasz}, {Kemmer}, {Kral}, {Lagrange}, {Launhardt}, {Matter},
  {M{\'e}nard}, {Milli}, {Molli{\`e}re}, {Olofsson}, {P{\'e}rez}, {Pinilla},
  {Pinte}, {Quanz}, {Schmidt}, {Udry}, {Wahhaj}, {Williams}, {Buenzli},
  {Cudel}, {Dominik}, {Galicher}, {Kasper}, {Lannier}, {Mesa}, {Mouillet},
  {Peretti}, {Perrot}, {Salter}, {Sissa}, {Wildi}, {Abe}, {Antichi},
  {Augereau}, {Baruffolo}, {Baudoz}, {Bazzon}, {Beuzit}, {Blanchard}, {Brems},
  {Buey}, {De Caprio}, {Carbillet}, {Carle}, {Cascone}, {Cheetham}, {Claudi},
  {Costille}, {Delboulb{\'e}}, {Dohlen}, {Fantinel}, {Feautrier}, {Fusco},
  {Giro}, {Gluck}, {Gry}, {Hubin}, {Hugot}, {Jaquet}, {Le Mignant}, {Llored},
  {Madec}, {Magnard}, {Martinez}, {Maurel}, {Meyer}, {M{\"o}ller-Nilsson},
  {Moulin}, {Mugnier}, {Orign{\'e}}, {Pavlov}, {Perret}, {Petit}, {Pragt},
  {Puget}, {Rabou}, {Ramos}, {Rigal}, {Rochat}, {Roelfsema}, {Rousset}, {Roux},
  {Salasnich}, {Sauvage}, {Sevin}, {Soenke}, {Stadler}, {Suarez}, {Turatto}, \&
  {Weber}}]{keppler2018}
{Keppler}, M., {Benisty}, M., {M{\"u}ller}, A., {et~al.} 2018, \aap, 617, A44

\bibitem[{Kim {et~al.}(2024)Kim, Fitzgerald, Lin, Sallum, Xin, Jovanovic, \&
  Leon-Saval}]{kim24interferometry}
Kim, Y.~J., Fitzgerald, M.~P., Lin, J., {et~al.} 2024, The Astrophysical
  Journal, 964, 113

\bibitem[{Kim {et~al.}(2022)Kim, Sallum, Lin, Xin, Norris, Betters, Leon-Saval,
  Lozi, Vievard, Gatkine, {et~al.}}]{kim2022spectroastrometry}
Kim, Y.~J., Sallum, S., Lin, J., {et~al.} 2022, in Ground-based and Airborne
  Instrumentation for Astronomy IX, Vol. 12184, SPIE, 1391--1402

\bibitem[{Kotani {et~al.}(2020)Kotani, Kawahara, Ishizuka, Jovanovic, Guyon,
  Vievard, Lozi, Sahoo, Yoneta, \& Tamura}]{kotani2020reach}
Kotani, T., Kawahara, H., Ishizuka, M., {et~al.} 2020, in Adaptive Optics
  Systems VII, Vol. 11448, International Society for Optics and Photonics,
  1144878

\bibitem[{Koza(2010)}]{koza2010sensitivity}
Koza, J. 2010, Solar Physics, 266, 261

\bibitem[{{Kuzuhara} {et~al.}(2022){Kuzuhara}, {Currie}, {Takarada}, {Brandt},
  {Sato}, {Uyama}, {Janson}, {Chilcote}, {Tobin}, {Lawson}, {Hori}, {Guyon},
  {Groff}, {Lozi}, {Vievard}, {Sahoo}, {Deo}, {Jovanovic}, {Ahn}, {Martinache},
  {Skaf}, {Akiyama}, {Norris}, {Bonnefoy}, {He{\l}miniak}, {Kudo}, {McElwain},
  {Samland}, {Wagner}, {Wisniewski}, {Knapp}, {Kwon}, {Nishikawa}, {Serabyn},
  {Hayashi}, \& {Tamura}}]{kuzuhara2022}
{Kuzuhara}, M., {Currie}, T., {Takarada}, T., {et~al.} 2022, \apjl, 934, L18

\bibitem[{Leon-Saval {et~al.}(2010)Leon-Saval, Argyros, \&
  Bland-Hawthorn}]{leon2010photonic}
Leon-Saval, S.~G., Argyros, A., \& Bland-Hawthorn, J. 2010, Optics Express, 18,
  8430

\bibitem[{Leon-Saval {et~al.}(2005)Leon-Saval, Birks, Bland-Hawthorn, \&
  Englund}]{leon2005multimode}
Leon-Saval, S.~G., Birks, T., Bland-Hawthorn, J., \& Englund, M. 2005, Optics
  letters, 30, 2545

\bibitem[{Levesque {et~al.}(2005)Levesque, Massey, Olsen, Plez, Josselin,
  Maeder, \& Meynet}]{levesque2005effective}
Levesque, E.~M., Massey, P., Olsen, K., {et~al.} 2005, The Astrophysical
  Journal, 628, 973

\bibitem[{{Levinstein} {et~al.}(2023){Levinstein}, {Sallum}, {Kim}, {Lin},
  {Lozi}, {Jovanovic}, {Fitzgerald}, \& {Vievard}}]{levinstein23}
{Levinstein}, D.~M., {Sallum}, S., {Kim}, Y.~J., {et~al.} 2023, in Society of
  Photo-Optical Instrumentation Engineers (SPIE) Conference Series, Vol. 12680,
  Society of Photo-Optical Instrumentation Engineers (SPIE) Conference Series,
  126800J

\bibitem[{Lin {et~al.}(2021)Lin, Jovanovic, \& Fitzgerald}]{lin2021design}
Lin, J., Jovanovic, N., \& Fitzgerald, M.~P. 2021, JOSA B, 38, A51

\bibitem[{Lin {et~al.}(2022)Lin, Vievard, Jovanovic, Norris, Fitzgerald,
  Betters, Gatkine, Guyon, Kim, Leon-Saval, {et~al.}}]{lin2022experimental}
Lin, J., Vievard, S., Jovanovic, N., {et~al.} 2022, in Advances in Optical and
  Mechanical Technologies for Telescopes and Instrumentation V, Vol. 12188,
  SPIE, 866--875

\bibitem[{Lozi {et~al.}(2019)Lozi, Jovanovic, Guyon, Chun, Jacobson, Goebel, \&
  Martinache}]{Lozi_2019}
Lozi, J., Jovanovic, N., Guyon, O., {et~al.} 2019, Publications of the
  Astronomical Society of the Pacific, 131, 044503

\bibitem[{Martinod {et~al.}(2021)Martinod, Norris, Tuthill, Lagadec, Jovanovic,
  Cvetojevic, Gross, Arriola, Gretzinger, Withford,
  {et~al.}}]{martinod2021scalable}
Martinod, M.-A., Norris, B., Tuthill, P., {et~al.} 2021, Nature communications,
  12, 2465

\bibitem[{Minowa {et~al.}(2010)Minowa, Hayano, Oya, Watanabe, Hattori, Guyon,
  Egner, Saito, Ito, Takami, {et~al.}}]{minowa2010performance}
Minowa, Y., Hayano, Y., Oya, S., {et~al.} 2010, in Adaptive Optics Systems II,
  Vol. 7736, International Society for Optics and Photonics, 77363N

\bibitem[{{Norris} {et~al.}(2022){Norris}, {Betters}, {Wei}, {Yerolatsitis},
  {Amezcua-Correa}, \& {Leon-Saval}}]{Norris2022}
{Norris}, B., {Betters}, C., {Wei}, J., {et~al.} 2022, Optics Express, 30,
  34908

\bibitem[{Norris {et~al.}(2020)Norris, Wei, Betters, Wong, \&
  Leon-Saval}]{norris2020all}
Norris, B.~R., Wei, J., Betters, C.~H., Wong, A., \& Leon-Saval, S.~G. 2020,
  Nature Communications, 11, 5335

\bibitem[{{Patat} {et~al.}(2007){Patat}, {Chandra}, {Chevalier}, {Justham},
  {Podsiadlowski}, {Wolf}, {Gal-Yam}, {Pasquini}, {Crawford}, {Mazzali},
  {Pauldrach}, {Nomoto}, {Benetti}, {Cappellaro}, {Elias-Rosa}, {Hillebrandt},
  {Leonard}, {Pastorello}, {Renzini}, {Sabbadin}, {Simon}, \&
  {Turatto}}]{patat07}
{Patat}, F., {Chandra}, P., {Chevalier}, R., {et~al.} 2007, Science, 317, 924

\bibitem[{{Pontoppidan} {et~al.}(2011){Pontoppidan}, {Blake}, \&
  {Smette}}]{pontoppidan11}
{Pontoppidan}, K.~M., {Blake}, G.~A., \& {Smette}, A. 2011, \apj, 733, 84

\bibitem[{{Reddy} {et~al.}(2022){Reddy}, {Topping}, {Shapley}, {Steidel},
  {Sanders}, {Du}, {Coil}, {Mobasher}, {Price}, \& {Shivaei}}]{reddy22}
{Reddy}, N.~A., {Topping}, M.~W., {Shapley}, A.~E., {et~al.} 2022, \apj, 926,
  31

\bibitem[{{Sallum} {et~al.}(2019){Sallum}, {Skemer}, {Eisner}, {van der Marel},
  {Sheehan}, {Close}, {Ireland}, {Males}, {Morzinski}, {Bailey}, {Briguglio},
  \& {Puglisi}}]{sallum19}
{Sallum}, S., {Skemer}, A.~J., {Eisner}, J.~A., {et~al.} 2019, \apj, 883, 100

\bibitem[{{Snellen} {et~al.}(2015){Snellen}, {de Kok}, {Birkby}, {Brandl},
  {Brogi}, {Keller}, {Kenworthy}, {Schwarz}, \& {Stuik}}]{snellen15}
{Snellen}, I., {de Kok}, R., {Birkby}, J.~L., {et~al.} 2015, \aap, 576, A59

\bibitem[{{Snellen} {et~al.}(2010){Snellen}, {de Kok}, {de Mooij}, \&
  {Albrecht}}]{snellen10}
{Snellen}, I. A.~G., {de Kok}, R.~J., {de Mooij}, E. J.~W., \& {Albrecht}, S.
  2010, \nat, 465, 1049

\bibitem[{Tinney {et~al.}(2004)Tinney, Ryder, Ellis, Churilov, Dawson, Smith,
  Waller, Whittard, Haynes, Lankshear, {et~al.}}]{tinney2004iris2}
Tinney, C.~G., Ryder, S.~D., Ellis, S.~C., {et~al.} 2004, in Ground-based
  Instrumentation for Astronomy, Vol. 5492, SPIE, 998--1009

\bibitem[{Tobin {et~al.}(2024)Tobin, Currie, Li, Chilcote, Brandt, Lacy,
  Kuzuhara, Vincent, El~Morsy, Deo, {et~al.}}]{tobin2024}
Tobin, T.~L., Currie, T., Li, Y., {et~al.} 2024, The Astronomical Journal, 167,
  205

\bibitem[{Vievard {et~al.}(2023{\natexlab{a}})Vievard, Huby, Lacour, Guyon,
  Cvetojevic, Jovanovic, Lozi, Barjot, Deo, Duchêne, Kotani, Marchis, Rouan,
  Martin, Lallement, Lapeyrere, Martinache, Ahn, Skaf, Tamura, Yuen, Lozi, \&
  Perrin}]{vievard2023singleaperture}
Vievard, S., Huby, E., Lacour, S., {et~al.} 2023{\natexlab{a}}, Single-aperture
  spectro-interferometry in the visible at the Subaru telescope with FIRST:
  First on-sky demonstration on Keho'oea (alpha Lyrae) and Hokulei (alpha
  Aurigae)

\bibitem[{Vievard {et~al.}(2023{\natexlab{b}})Vievard, Lallement, Huby, Lacour,
  Guyon, Jovanovic, Leon-Saval, Lozi, Deo, Ahn, {et~al.}}]{vievard2023photonic}
Vievard, S., Lallement, M., Huby, E., {et~al.} 2023{\natexlab{b}}, in
  Techniques and Instrumentation for Detection of Exoplanets XI, Vol. 12680,
  SPIE, 137--153

\bibitem[{Vigan {et~al.}(2024)Vigan, El~Morsy, Lopez, Otten, Garcia, Costes,
  Muslimov, Viret, Charles, Zins, {et~al.}}]{vigan2024first}
Vigan, A., El~Morsy, M., Lopez, M., {et~al.} 2024, Astronomy \& Astrophysics,
  682, A16

\bibitem[{{Wagner} {et~al.}(2023){Wagner}, {Stone}, {Skemer}, {Ertel}, {Dong},
  {Apai}, {Spalding}, {Leisenring}, {Sitko}, {Kratter}, {Barman}, {Marley},
  {Miles}, {Boccaletti}, {Assani}, {Bayyari}, {Uyama}, {Woodward}, {Hinz},
  {Briesemeister}, {Lawson}, {M{\'e}nard}, {Pantin}, {Russell}, {Skrutskie}, \&
  {Wisniewski}}]{wagner23}
{Wagner}, K., {Stone}, J., {Skemer}, A., {et~al.} 2023, Nature Astronomy, 7,
  1208

\bibitem[{Xin {et~al.}(2022)Xin, Jovanovic, Ruane, Mawet, Fitzgerald,
  Echeverri, Lin, Leon-Saval, Gatkine, Kim, {et~al.}}]{xin2022efficient}
Xin, Y., Jovanovic, N., Ruane, G., {et~al.} 2022, The Astrophysical Journal,
  938, 140

\bibitem[{Young {et~al.}(2003)Young, Baldwin, Basden, Bharmal, Buscher, George,
  Haniff, Keen, O'Donovan, Pearson, Thorsteinsson, Thureau, Tubbs, \&
  Warner}]{10.1117/12.459135}
Young, J.~S., Baldwin, J.~E., Basden, A.~G., {et~al.} 2003, in Interferometry
  for Optical Astronomy II, ed. W.~A. Traub, Vol. 4838, International Society
  for Optics and Photonics (SPIE), 369 -- 378

\bibitem[{{Young} {et~al.}(2000){Young}, {Baldwin}, {Boysen}, {Haniff},
  {Lawson}, {Mackay}, {Pearson}, {Rogers}, {St. -Jacques}, {Warner}, {Wilson},
  \& {Wilson}}]{2000MNRAS.315..635Y}
{Young}, J.~S., {Baldwin}, J.~E., {Boysen}, R.~C., {et~al.} 2000, \mnras, 315,
  635

\end{thebibliography}

\appendix

\section{Optical setup parameters for various numerical apertures of the beam injected into the PL}
\label{Appendix-optical-setup}

\noindent Table~\ref{tab:OpticalSetUp} provides optimized parameters of the PL injection bench for reference. Parameters are given for various focal ratios of the beam injected into the PL. Distances between the optical components are measured from the back surface of the first optical component to the front surface of the second optical component. For focal ratios of $10.6$, $12$, and $28$, only two lenses are needed (L1 and L2). The back focal length provided in the table corresponds to the distance between the back surface of the last lens (L2 or L3, depending on the focal ratio considered) and the focal plane.

\begin{table}[h!]
    \centering
    \caption{Optical setup of the PL injection.}
    \begin{tabular}{|c|c|c|c|c|c|c|}
    \hline
   Focal ratio & 6 & 8 & 10.6 & 12 & 28 \\
   \hline 
    BS to L1 & 62.6 & 335.2 & 242.2 & 183.4 & 244.6 \\
   \hline       
    \makecell{L1 focal \\ length} & 500 & 500 & 400 & 400 & 400 \\
   \hline       
    L1 to L2 & 134.1 & 37.2 & 207.1 & 241.6 & 10 \\
   \hline       
    \makecell{L2 focal \\ length} & 400 & 400 & 150 & 150 & 400 \\
   \hline       
    L2 to L3 & 332.2 & 119.6 & 20.7 & 45 & 210 \\
   \hline       
 \makecell{L3 focal \\ length} & 50 & 150 & N/A & N/A & N/A \\
   \hline       
   \makecell{Back focal \\ length} & 43.0 & 91.7 & 124.4 & 109.4 & 181.2\\
   \hline
   \makecell{Effective \\ focal length} & -134.2 & 140.4 & 171.2 & 189.9 & 199.6 \\
   \hline 
\end{tabular}
    \tablefoot{Optical setup parameters for various numerical apertures of the beam injected into the PL. Distances and focal lengths are given in mm. (N/A\,: Not Applicable)}
    \label{tab:OpticalSetUp}
\end{table}



\end{document}